\documentclass[aps,twocolumn]{revtex4-1}
\usepackage{amsmath,amssymb}
\usepackage{slashed}
\usepackage{graphicx}
\usepackage{bm}
\usepackage[T1]{fontenc}
\usepackage[utf8]{inputenc}
\usepackage{gauss} 
\usepackage[top=1.0in,bottom=1.0in,left=1.0in,right=1.0in]{geometry}
\usepackage[colorlinks,linkcolor=blue,citecolor=blue]{hyperref}
\usepackage{comment}
\usepackage{ytableau}
\newcommand{\be}{\begin{equation}}
\newcommand{\ee}{\end{equation}}
\newcommand{\bea}{\begin{eqnarray}}
\newcommand{\eea}{\end{eqnarray}}
\newcommand{\ba}{\begin{eqnarray}}
\newcommand{\ea}{\end{eqnarray}}

\newcommand{\ket}[1]{\left|#1\right\rangle}

\begin{document}

\title{Glueballs, Constituent Gluons and Instantons}

\author{Edward Shuryak}
\email{edward.shuryak@stonybrook.edu}
\affiliation{Center for Nuclear Theory, Department of Physics and Astronomy, Stony Brook University, Stony Brook, New York 11794-3800, USA}

\author{Ismail Zahed}
\email{ismail.zahed@stonybrook.edu}
\affiliation{Center for Nuclear Theory, Department of Physics and Astronomy, Stony Brook University, Stony Brook, New York 11794-3800, USA}

\begin{abstract}
We present a constituent two-gluon description of the lowest-lying glueball states in pure Yang--Mills theory, calibrated against quenched lattice results. The framework incorporates an instanton-induced dynamical gluon mass, Casimir-scaled adjoint confinement, the short-distance adjoint Coulomb interaction, and instanton-induced central and tensor forces. 
The scalar $0^{++}$ glueball is found to be exceptionally compact, with a radius of order the instanton size, $\rho \sim \frac 13\,\mathrm{fm}$, consistent with lattice indications. By contrast, the tensor $2^{++}$ state remains spatially extended due to the centrifugal barrier. We also discuss the role of $S$--$D$ mixing. %A two-gluon vector glueball is ruled out by the Landau--Yang selection rule.
A semiclassical analysis further supports Regge behavior for excited states, in agreement with lattice results.
\end{abstract}

%\date{}

\maketitle

\section{Introduction}
\subsection{Historic remarks}
Glueballs are the  color-singlet bound states of non-Abelian gauge theory, composed entirely of gluonic degrees of freedom. Their
comparison with mesons
provide a unique window into mechanism of confinement, dynamical mass generation, and the role of topology in quantum chromodynamics (QCD).

One of the most obvious question in QCD is why the observed hadrons are made of quarks and not of gluons. It appears that glueballs are much heavier than typical quark- model hadrons, and therefore they have large widths and/or complicated decay patterns, making them difficult to find. But why are glueballs so heavy? 
What are their masses, radii and other parameters in a purely gluonic world, and how do they change if one includes light quarks?

In the pure Yang--Mills (``quenched'') theory, where quark degrees of freedom are absent, the glueball spectrum has been extensively determined using lattice gauge theory, which remains the most reliable nonperturbative approach to this problem \cite{Morningstar:1999rf,Meyer:2003rf,Chen:2006dv,Athenodorou:2020ani}. 
Early high-precision lattice calculations, particularly those employing anisotropic lattices, established the ordering of the lowest-lying glueball states and provided quantitative benchmarks for their masses. These studies identified the scalar $0^{++}$ glueball as the lightest state, with mass $m_{0^{++}} \approx 1.6\,\mathrm{GeV}$, followed by the tensor $2^{++}$ state with $m_{2^{++}} \approx 2.2\,\mathrm{GeV}$. The pseudoscalar $0^{-+}$ state appears at $m_{0^{-+}} \approx 2.4\,\mathrm{GeV}$, with higher-spin excitations occurring at significantly larger masses \cite{Morningstar:1999rf,Chen:2006dv}. 
Subsequent studies employing improved actions, enlarged operator bases, and controlled continuum extrapolations have refined this picture and confirmed its robustness across lattice spacings and volumes. In particular, recent high-statistics calculations have provided a precise reference spectrum for pure $\mathrm{SU}(3)$ Yang--Mills theory, covering a wide range of $J^{PC}$ channels \cite{Athenodorou:2020ani}.

While the quenched glueball spectrum is now comparatively well established, several important questions have gained renewed attention in recent years. One concerns the {\em internal structure} of glueballs in hadrons, including their spatial extent and form factors. Lattice studies of energy-momentum tensor matrix elements and related gravitational form factors have begun to probe the size and mass distribution of glueballs \cite{Meyer:2010ku,abbott2025gravitationalformfactorsglueballs}, with emerging evidence that the scalar $0^{++}$ state may be unusually compact compared to higher-spin excitations. Such results provide new, nontrivial constraints on phenomenological and microscopic models of glueballs. A recent review summarizes the current status of lattice calculations, experimental searches, and theoretical interpretations \cite{Morningstar:2025glueballreview}.

\begin{figure}
    \centering
\includegraphics[width=0.75\linewidth]{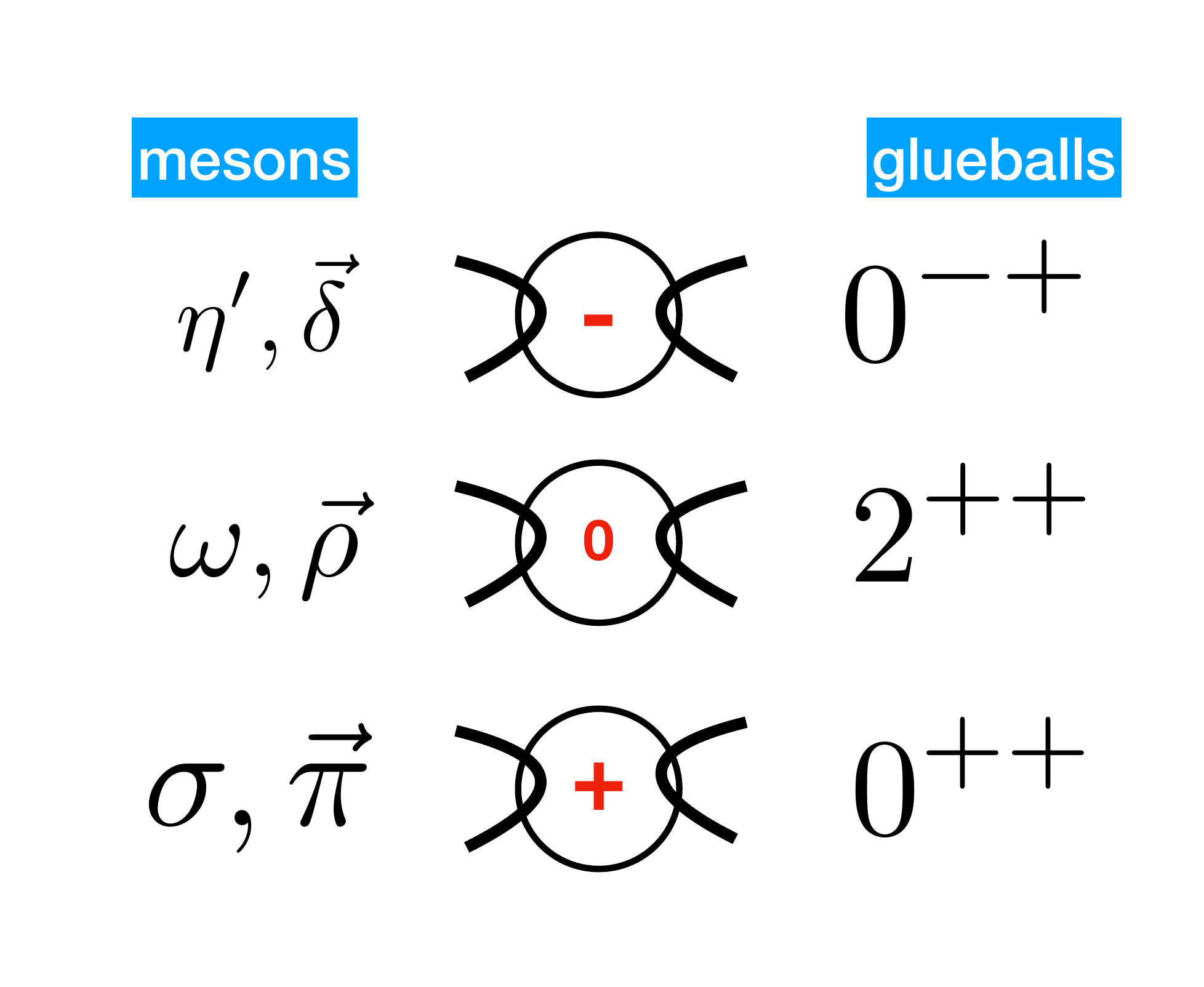}
    \caption{Repulsive, neutral and attractive channels induced by instanton-induced effects in Euclidean correlation functions, for mesons and glueballs. }
    \label{fig_inst_channels}
\end{figure}

The nonperturbative interactions in the scalar, tensor, and pseudoscalar correlation functions at short distances were related, already three decades ago, to instanton-induced forces \cite{Schafer:1994fd,Schafer:1996wv}; see Fig.~\ref{fig_inst_channels}. 
In particular, these studies showed that instantons generate a strong attractive interaction in the scalar ($J^{PC}=0^{++}$) channel. As a consequence, it was predicted that the scalar glueball should be significantly more compact than typical hadrons, with a mean square radius $r_{\mathrm{m.s.}} \approx 0.2\,\mathrm{fm}$. 
By contrast, the instanton-induced interaction is repulsive in the pseudoscalar ($0^{-+}$) channel, while it is absent in the tensor ($2^{++}$) channel. Consequently, these three glueball channels exhibit a pattern analogous to that of the $\sigma$ (scalar), $\eta'$ (pseudoscalar), and $\rho,\omega$ (vector) mesons, respectively.

\begin{figure}[h]
\centering
\includegraphics[width=0.65\linewidth]{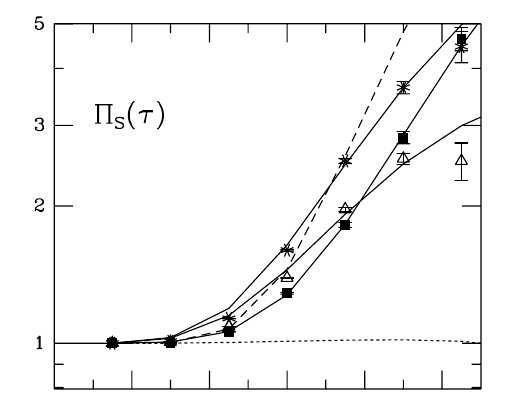}
\includegraphics[width=0.65\linewidth]{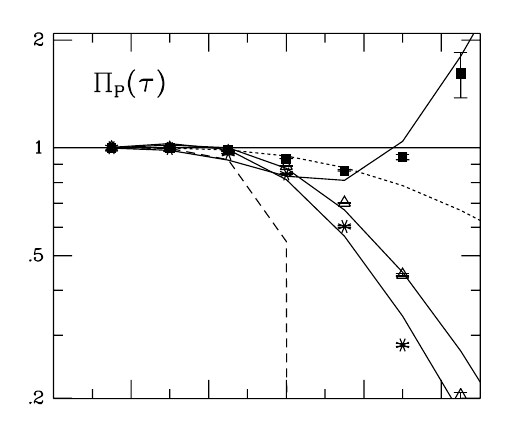}
\caption{Scalar $G^2$ and pseudoscalar $G\tilde G$ gluonic correlation functions
normalized to the corresponding free correlators as functions of the
Euclidean time separation, taken from~\cite{Schafer:1994fd}. Results in
the random, quenched, and full ensembles are denoted by stars, open
triangles, and solid squares, respectively. The solid lines show the
parametrization described in the text, the dashed line the dilute
instanton gas approximation, and the dotted line the QCD sum-rule
calculation. The horizontal line in the second panel is added to guide
the eye. }
\label{fig_corr}
\end{figure}

%%%%%%%
Correlation functions of scalar, pseudoscalar, and tensor glueball
operators calculated in the instanton liquid model (ILM)
go back to \cite{Schafer:1994fd} are reproduced in Fig.~\ref{fig_corr}. They
clearly show that the interaction in these three channels is attractive,
repulsive, and weak, respectively.

 Values for the masses and radii of the lowest glueball states,
extracted from fits to the correlation functions and Bethe-Salpeter
amplitudes 30 years ago in \cite{Schafer:1994fd}, are listed in Table~\ref{tab_corr}.

\begin{table}[h]
\centering
\begin{tabular}{|c|c|c|c|} \hline
 & scalar & pseudoscalar & tensor \\ \hline
 $M$ (GeV) & $1.4\pm0.2$ & $>3$ & ? \\ \hline
 $r_{\rm r.m.s.}$ (fm) & 0.21 & ? & 0.61 \\ \hline
\end{tabular}
\caption{Masses and sizes of glueballs from fits to correlation functions
and Bethe-Salpeter amplitudes, as predicted in~\cite{Schafer:1994fd}.}
\label{tab_corr}
\end{table}

%%%%%%%

Let us now briefly review the current experimental status of these three glueballs. The scalar glueball mass, around $m_{0^{++}}\approx 1.6\, \text{GeV}$, lies in a region populated by several scalar mesons. Its admixture with them remains a subject of ongoing investigation, with perhaps the largest share residing in $f_0(1710)$.
The tensor state $M_{2^{++}}\approx 2.2\, \text{GeV}$ has perhaps been observed in ``glue-rich'' double-diffractive scattering processes.

Interestingly, the pseudoscalar glueball has received a boost following the BESIII announcement of the discovery of the pseudoscalar $X(2370)$ resonance, with mass and width $m=2395\, \text{MeV}$ and $\Gamma=188 \, \text{MeV}$, which fit well with expectations for the pseudoscalar glueball. The reaction is
$$ J/\psi \rightarrow \gamma g g, \,\, gg\rightarrow X(2370) \rightarrow \eta' f_0(980) $$

``Holographic QCD'' is a theoretical framework that combines mesons and glueballs within a common geometrical construction, see e.g.\cite{Gursoy:2007cb}. It has significant predictive power: for example, the complicated meson--glueball mixing has been studied using this model in \cite{Iatrakis:2015rga}. (This mixing defines the mutual interaction strength of the QCD flux tubes, which is important for event generators at colliders.)

Mixing of glueballs with mesons is, of course, also investigated on the lattice in full QCD. When light quarks are dynamical, pure-glue operators can mix with flavor-singlet $q\bar q$ states, particularly in the scalar and pseudoscalar channels. This mixing with light mesons complicates the identification of experimental candidates and blurs the connection between quenched lattice spectra and observed isoscalar mesons.

Lattice calculations incorporating dynamical quarks have begun to address these issues directly, including studies of pseudoscalar glueball--$\eta$ and $\eta'$ mixing \cite{Jiang:2022ffl}.

In parallel with lattice developments, a variety of theoretical approaches have been pursued. Functional methods based on Dyson--Schwinger and Bethe--Salpeter equations reproduce many qualitative features of the glueball spectrum \cite{Alkofer:2004it,Fischer:2012ty,Holler:2015bua}. Phenomenological models, including constituent-gluon Hamiltonians \cite{Cornwall:1982zn}, flux-tube \cite{Isgur:1984bm} and bag models \cite{Johnson:1975sg}, as well as holographic constructions \cite{Brunner:2015oqa}, provide intuitive pictures for the organization of glueball states and their Regge behavior.

We have already mentioned instanton-based contributions to Euclidean correlation functions \cite{Schafer:1996wv}, related to glueball masses and couplings via certain sum rules.

In this paper, however, we follow a different path, based on the notion of ``effective gluons,'' with a momentum-dependent dynamical gluon mass that induces channel-dependent short-range interactions \cite{Musakhanov:2017erp,Shuryak:2021xkq}. Together with well-separated instantons, we also include contributions from instanton--antiinstanton ``molecules.'' While this approach also generates enhanced attraction for the $0^{++}$ glueball at small distances, its parametric strength (in this and other channels) turns out to be different.
 
\subsection{Outline of this work}

Having completed this brief overview, let us now explain the motivations of the present work. Note that two (out of three) glueballs mentioned -- $0^{++},2^{++},0^{-+}$ -- are ``exceptional,'' in the same sense as their mesonic analogues, $\sigma,\rho,\eta'$. In fact, only vector mesons are ``normal,'' in the sense that, for example, one can predict baryon masses as $M_{baryons}\approx (3/2)M_{mesons}$ using a simple additive count in a ``constituent'' model, in which the mean ``interaction'' part of the Hamiltonian is subleading due to partial cancellations.

What we attempt below is to construct a constituent two-gluon description of glueball states. In doing so, we do not focus on the few ``exceptional'' states discussed above, but instead first aim to reproduce the ``bulk'' of glueball spectroscopy. Indeed, if quark models of mesons and baryons were to start with $\pi,\eta'$, they would not be successful either.

Our framework combines adjoint Coulomb interactions fixed by group theory, Casimir-scaled confinement with gluonic screening. Only after this is established do we add instanton-induced central and spin-dependent forces.

Further motivation for this work comes from our broader project of ``bridging hadron spectroscopy with partonic observables,'' which aims to translate quark models of mesons, baryons, and multiquark states from the rest frame to their corresponding light-front formulations. For completeness, it is clearly necessary to include glueballs and quark--gluon hybrids in the Fock components of hadrons. The renormalization group, in the form of an expanding set of admixed states in hadronic wave functions, would then replace the current ``evolution equations,'' which are defined for PDFs rather than wave functions.

\begin{figure}[h!]
    \centering    \includegraphics[width=0.95\linewidth]{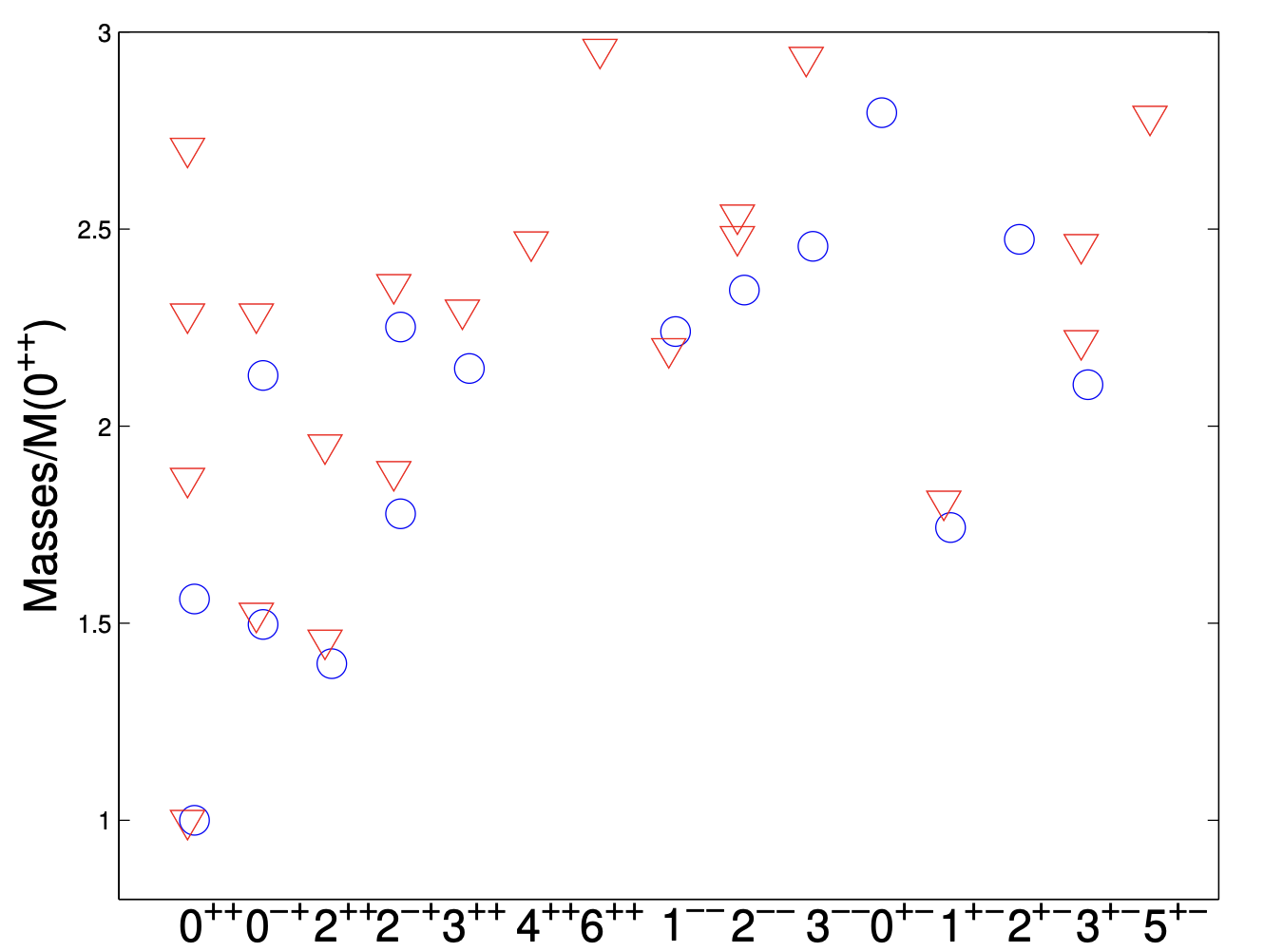}
    \caption{Lattice data for glueball masses, normalized to lowest scalar,  triangles from Meyer et al and  circles from Morningstar et al.} \label{fig_lattice_glueballs}
\end{figure}

To convince the reader that there exists a large set of glueball states with various quantum numbers, we reproduce Fig.\ref{fig_lattice_glueballs} from \cite{Mathieu:2008me}, comparing results from two different lattice groups. Not only are there many states in total, but in the scalar case the celebrated ground state is supplemented by three excited states, and in several channels two states have been identified.

We first focus on a subset of excited states, shown in Table~\ref{tab_corr}. A surprising outcome of fitting these states is the rather heavy ``constituent gluon mass,'' $m_g\sim 900\, \text{MeV}$, well above typical values in the literature. Only later do we return to the ``exceptional'' lowest states in the scalar and pseudoscalar sectors.

%\begin{widetext}
%\end{widetext}

\begin{table}[h!]
    \centering
    \begin{tabular}{|c|c|c|c|} \hline
J= 0 &  1.475 &(30) &(65) \\
  0 & 2.755 &(70) &(120)  \\
  0 & 3.370 & (100)& (150) \\
   0& 3.990 &(210) &(180) \\
  2 & 2.150 &(30) &(100)  \\
  2 & 2.880 &(100) &(130) \\
  3 & 3.385 & (90) &(150) \\
  4 & 3.640 &(90) &(160)  \\
  6 & 4.360 & (260) &(200) \\ \hline
 \end{tabular}
    \caption{Energies (GeV) and error bars of all glueballs, for all $P=C=+$ states, from \cite{Meyer:2004gx}}
    \label{tab_all_pp}
\end{table}

What we find is that these states can, in fact, be described by a Schr\"odinger equation without invoking any extreme assumptions. Moreover, the spectrum of ``normal'' glueballs turns out to lie between that of strange $\bar s s$ and charmed $\bar c c$ mesons \cite{Cornwall:1982zr,Mathieu:2008me}.

Particular emphasis is placed on the scalar and tensor channels. While the $0^{++}$ state is dominated by short-distance dynamics and can become relatively compact, the $2^{++}$ glueball is affected by angular momentum barriers and significant $S$--$D$ wave mixing. We solve the resulting coupled-channel problem and discuss the mass hierarchies and spatial structure.

%%%%%%%%%%%%%%%%%%%%%%%%%%%%%%%%%%%%%%%%%%%%%%%%%%

The paper is organized as follows. In Sec.~\ref{sec:hamiltonian} we construct the constituent two-gluon Hamiltonian, including adjoint Coulomb interaction, Casimir-scaled confinement with screening, and short-distance nonperturbative forces induced by instantons. In Sec.~\ref{sec:schrodinger} we solve the resulting Schr\"odinger equation and establish the spectrum of ``normal'' glueballs, focusing on radial and orbital excitations and their comparison to lattice results. The lowest scalar and pseudoscalar channels, where short-distance effects are dominant, are analyzed separately, including a semiclassical WKB baseline.

In Sec.~\ref{sec:relativistic} we develop a relativistic treatment of the scalar $0^{++}$ glueball based on an instanton-induced four-gluon interaction and its reduction to an effective two-body kernel. This leads to a reduced Bethe--Salpeter (Salpeter) equation, from which we extract the mass and wavefunction of the compact scalar state. In Sec.~\ref{subsec:spectrum_PC} we assemble the full glueball spectrum, highlighting the interplay between confinement-driven Regge behavior and channel-dependent instanton dynamics across different $J^{PC}$ sectors. Our conclusions are summarized in Sec.~\ref{sec_summary}.

Technical details and complementary derivations are collected in the appendices. These include the WKB quantization procedure and semiclassical estimates, the construction and properties of the screened adjoint potential, details of instanton-induced interactions and spin-dependent forces, the derivation of the effective four-gluon vertex and its reduction to a two-body interaction, and the helicity and partial-wave projections relevant for the scalar and tensor channels.

\section{The Hamiltonian}
\label{sec:hamiltonian}

\subsection{Adjoint Coulomb}
The Coulomb term follows from one-gluon exchange. Writing
\begin{equation}
V_C(r)=\frac{\alpha_s}{r}\,T_1\!\cdot T_2,
\qquad
T_1\!\cdot T_2=\frac12\big(C_R-C_{R_1}-C_{R_2}\big),
\label{eq:VCgeneral}
\end{equation}
one obtains the Casimir form
\begin{equation}
V_C(r)=-\frac{\alpha_s}{2r}\big(C_{R_1}+C_{R_2}-C_R\big).
\label{eq:VCcasimir}
\end{equation}
For two gluons with $R_1=R_2=A$ and $C_A=3$, coupled to a singlet $R=1$ with
$C_R=0$, this reduces to
\begin{equation}
V_C^{(gg,1)}(r)=-\frac{3\alpha_s^{\rm eff}}{r},
\label{eq:VCgg}
\end{equation}
which is attractive and $9/4$ times stronger than the $q\bar q$ singlet
Coulomb coefficient at the same value of $\alpha_s^{\rm eff}$.
We will meet rescaling by $9/4$
in other quantities, see below.

\subsection{Adjoint confining potential}
We model glueballs as two effective constituent gluons in their center-of-mass frame. The relative motion Hamiltonian is
\begin{eqnarray}
&&H=2m_g(\eta)+\frac{\bm p^2}{m_g(\eta)} +V_0\nonumber\\
&&+V_{\rm conf}(r)+V_C(r)+V_{\rm inst}(r)+V_{SS}(r)+V_T(r).\nonumber\\
\label{eq:Hfull}
\end{eqnarray}

The confining term is taken as a screened adjoint (double) string,
\begin{eqnarray}
V_{\rm conf}(r)&=&\sigma_8\,r\,e^{-r/r_{\rm scr}}+V_0,
\nonumber\\
\sigma_8&=&\frac{C_A}{C_F}\sigma_3=\frac94\,\sigma_3,
\nonumber\\
\sigma_3&=&0.18~\mathrm{GeV}^2,
\label{eq:Vconf}
\end{eqnarray}
The overall shift by a constant $V_0$
is not known \emph{a priori}, but fits to the spectrum
yield a rather modest negative value. This shift can
be avoided altogether if only level spacings, rather
than absolute masses, are used in the fits.

\begin{figure}
    \centering  \includegraphics[width=0.75\linewidth]{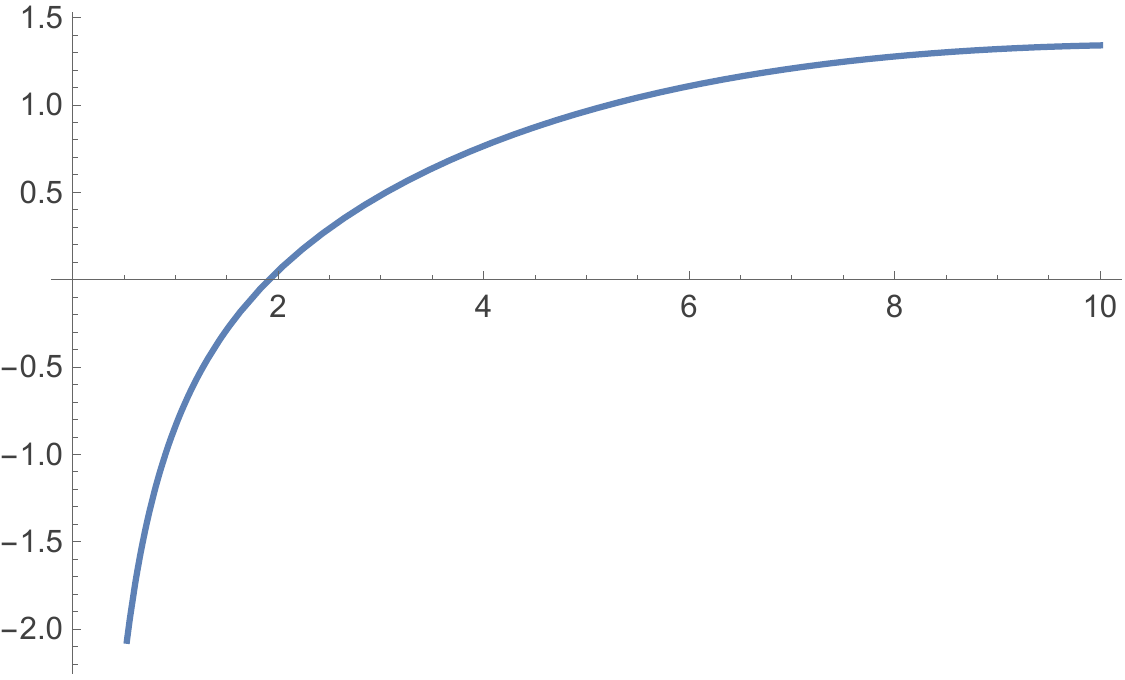}
    \caption{Central potential $V_{conf}(r)\, (\rm GeV)$ used versus $r \,(\rm GeV^{-1})$}
    \label{fig_pot}
\end{figure}

The choice of the adjoint screening length $r_{\rm scr}$
is a rather delicate issue. If $r_{\rm scr}$ is smaller
than the typical size of the states, the potential
acts merely as a barrier and is unable to support
bound states of larger spatial extent, producing at
most a few quasibound states, if any. To avoid this
situation we place the states in a box of radius
$R = 2\,\mathrm{fm}$ and take $r_{\rm scr}\sim R$, so that
the potential inside the box does not decrease but
instead saturates, as illustrated in
Fig.~\ref{fig_pot}. A more detailed discussion, as
well as the relation of this potential shape to our
Monte-Carlo calculations with a single instanton or
$\bar I I$ molecule using numerically generated Wilson
lines, is given in Appendix~\ref{sec_pote_numeric}.

\subsection{The short-distance nonperturbative interactions}

Let us start with simple estimates
of instanton-induced interactions. Naively representing the gauge field as
semiclassical O(1/g) plus quantum
fluctuations O(1)
$A=A_{\rm class}+A_{\rm gluon}$ yields a cross term of order $O(1/g)$. Yet it
 vanishes upon averaging over random instanton orientations. A
correct estimate instead follows from the quartic term
\[
g^2 A_{\rm class}^2 A_{\rm gluon}^2 \sim O(g^0)\,A_{\rm gluon}^2,
\]
which contributes to the effective gluon mass. Note that for light
quarks the instanton-induced effective Lagrangian is built from fermion
zero modes, and likewise carries no explicit factor of $g$. 
So, parametrically we expect constituent masses be similar.

Numerically,
however, in glueballs this interaction is strong due to
color Casimir factors. This explains why the scalar
glueball is predicted to be unusually compact, with a size
$\sim 0.2\,\mathrm{fm}$, whereas even the smallest meson, the pion, has a
radius of order $\sim 0.5\,\mathrm{fm}$.

The perturbative spin-spin interaction is local
$\sim\delta^{(3)}(r)$. The instanton-induced attraction is modeled by a
Gaussian centered at the origin with range $\rho$,
\begin{equation}
V_{\rm inst}(r)=-G(\eta)\,e^{-r^2/2\rho^2},
\qquad
G(\eta)=G_0\,\eta,
\label{eq:Vinst}
\end{equation}
with $\rho\simeq \frac13\,\mathrm{fm}$.  Because the (anti)instanton field is
(anti)self-dual, the induced coupling derived in~\cite{Liu:2024glue}
(see Eq.~(45) therein) involves the ’t~Hooft symbols $\eta,\bar\eta$ and
projects most strongly onto the parity-even scalar combination
$G^a_{\mu\nu}G^a_{\mu\nu}$, enhancing the $0^{++}$ channel, while the
Levi-Civita structure governs the pseudoscalar $G\tilde G$ channel.

The spin-dependent interactions are taken in the standard form
\begin{eqnarray}
V_{SS}(r) &=&
\frac{C_{SS}(\eta)}{m_g^2(\eta)}\,
\delta^{(3)}_\Lambda(\bm r)\,\bm S_1\!\cdot\!\bm S_2,
\nonumber\\
V_T(r) &=&
\frac{C_T(\eta)}{m_g^2(\eta)}\,
\frac{1-e^{-r^2/\rho^2}}{r^3}\,S_{12},
\label{eq:Vsd}
\end{eqnarray}
with
$S_{12}=3(\bm S_1\!\cdot\!\hat{\bm r})(\bm S_2\!\cdot\!\hat{\bm r})
-\bm S_1\!\cdot\!\bm S_2$, and a Gaussian-smeared contact regulator
\begin{equation}
\delta^{(3)}_\Lambda(\bm r)
=
\left(\frac{\Lambda^2}{\pi}\right)^{3/2}
e^{-\Lambda^2 r^2},
\qquad
\Lambda\sim\rho^{-1}.
\label{eq:deltaReg}
\end{equation}
Dense-ILM including instanton-antiinstanton correlated pairs or ``molecules" use parameter $\eta\sim 7.$ for their effects, as compared to dilute-ILM with only
instantons, and density $n_{dilute}\approx 1\, fm^{-4}$
This
scaling is implemented through
\begin{equation}
C_{SS}(\eta)=C_{SS}^{(0)}\,\eta,
\qquad
C_T(\eta)=C_T^{(0)}\,\eta,
\label{eq:Ceta}
\end{equation}
with $C_T^{(0)}$ allowed to be negative, consistent with the ILM result
of~\cite{Shuryak:2021fsu}. The dynamical constituent gluon mass is
parametrized as
\begin{equation}
m_g(\eta)=m_{g0}\sqrt{\eta},
\qquad
m_{g0}\simeq 0.36~\mathrm{GeV}.
\label{eq:mgEta}
\end{equation}

\section{Glueball spectra from the Schr\"odinger equation}
\label{sec:schrodinger}
\subsection{The $S=0,\,J=L$ sector}

As already mentioned, since the effective gluon mass is
$O(1\,\mathrm{GeV})$, the accuracy of the nonrelativistic approximation
is expected to be comparable to that for charmonium. Of course, important differences arising from the distinct coefficients in the Hamiltonians.

\begin{figure}
    \centering
    \includegraphics[width=0.85\linewidth]{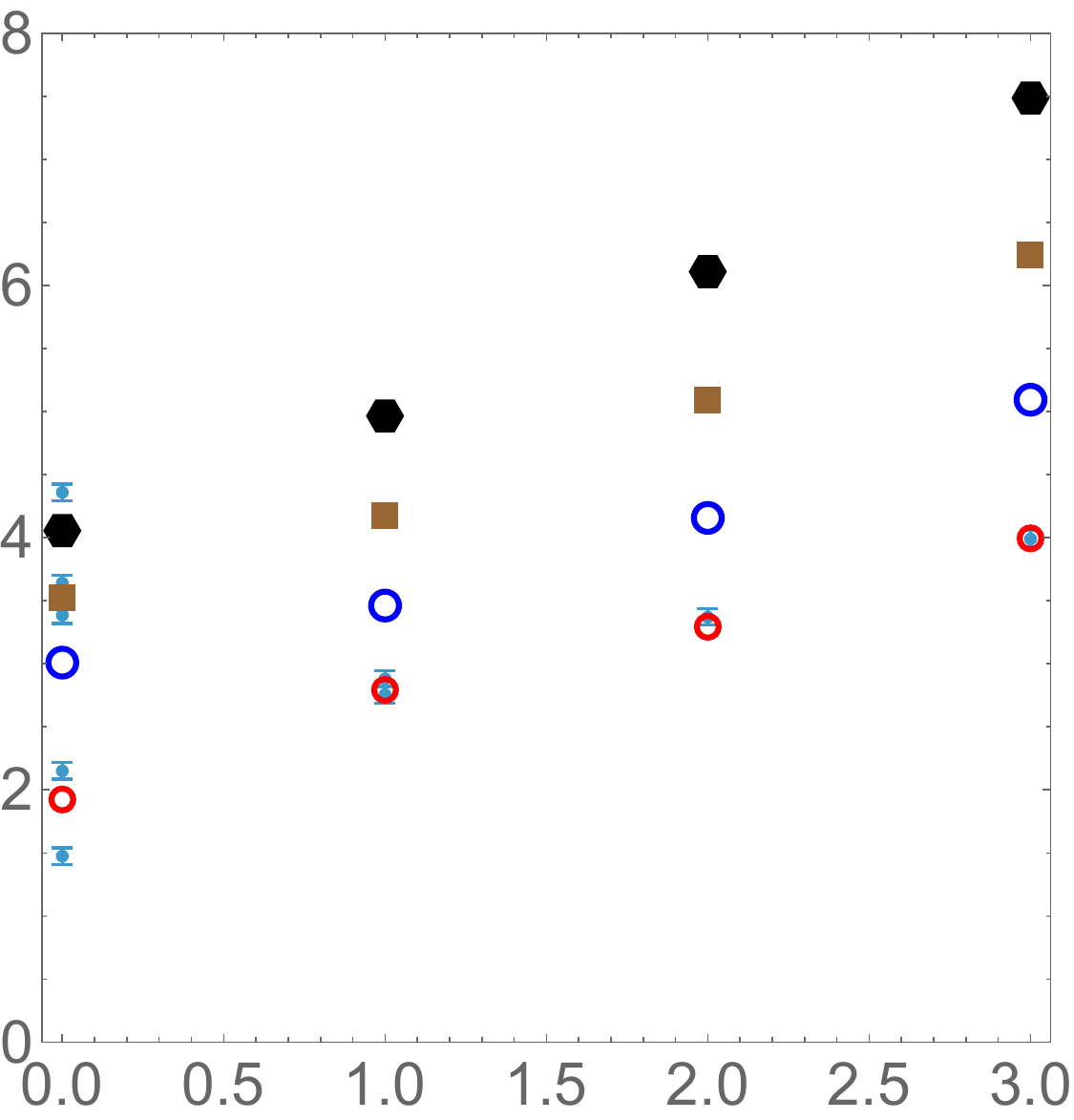}
    \caption{The colored points show the calculated energies $E_0,E_1,E_2,E_3$ (GeV) of the
    four lowest states 
    with
    $L=0,2,4,6$.
Small points with error bars are H.Meyer's 
lattice simulations in pure SU(3) gauge
    theory \cite{Meyer:2004gx}.
    We used $J^{PC}$ lattice states
    $0^{++}$, $2^{++}$, $4^{++}$, and $6^{++}$ (from bottom
    to top) as a function of the principal quantum number $n$.  }
    \label{fig_gbpp}
\end{figure}

We begin with the simplest case, with total spin $S=0$, so that the total
and orbital angular momenta coincide, $J=L$. The calculated spectrum of
several $J^{PC}=J^{++}$ channels is shown in
Fig.~\ref{fig_gbpp}. At this stage only the Coulomb and confining
potentials are included, with no spin-dependent forces, as is commonly
done, for example, in charmonium studies. The model parameters are
chosen to reproduce the three higher states in the scalar channel,
which, unlike the lowest $n=0$ state, have ``normal'' hadronic sizes
(see the table of r.m.s.\ radii) and are therefore less sensitive to
short-distance forces. The most important parameter obtained in this
way is the effective gluon mass, fitted to be
\be
m_g = 0.90\,\mathrm{GeV}.
\ee
For comparison, the effective masses of the strange and charmed quarks
are $m_s \approx 0.5\,\mathrm{GeV}$ and $m_c \approx 1.5\,\mathrm{GeV}$,
respectively.

This implies a value of the parameter $\eta$—the enhancement of the
gluon mass relative to the dilute ILM value due to instanton–antiinstanton
molecules—of $\eta = 6.25$. This is close to the value $\eta \approx 7$
used in \cite{Shuryak:2021fsu} to reproduce the central charmonium
potential. That value was originally motivated by ``gradient flow''
cooling of lattice gauge configurations \cite{Shuryak:2021fsu}.

\begin{figure}
    \centering
    \includegraphics[width=0.85\linewidth]{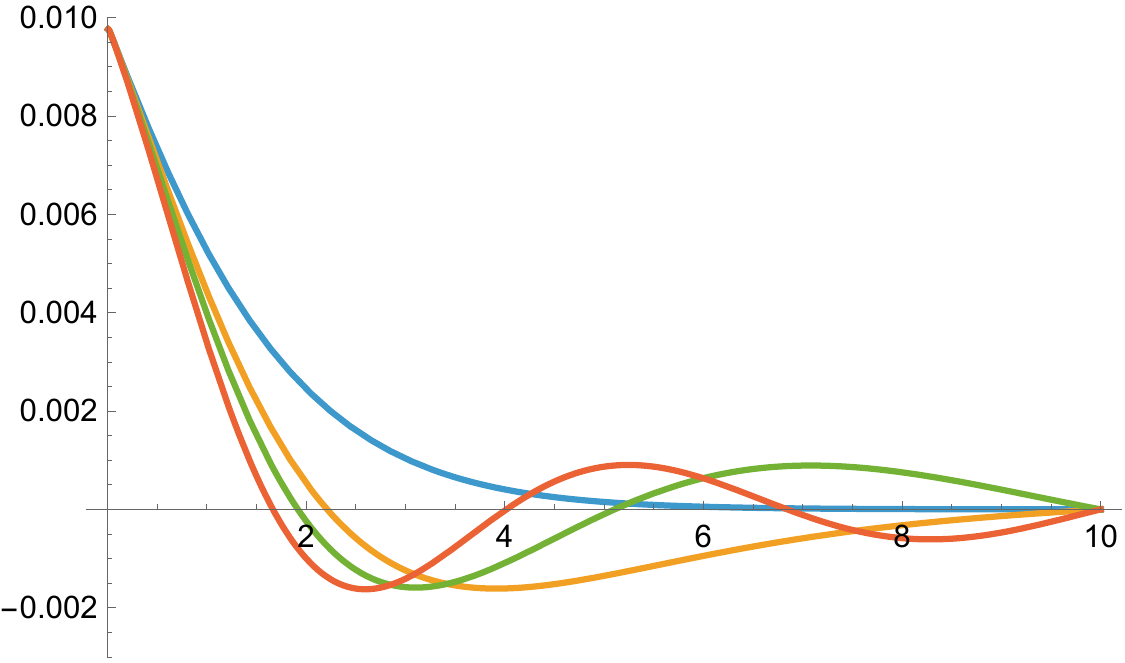}
    \caption{Wave functions (unnormalized) $\psi_n(r)$ for $n=0,1,2,3$ as
    functions of $r$ ($\mathrm{GeV}^{-1}$).}
    \label{fig_GB-func}
\end{figure}

To quantify the sizes and shapes of these states, we display the
corresponding wave functions in Fig.~\ref{fig_GB-func} and list the
root-mean-square radii,
\be
r_{\rm r.m.s.}
=
\left[
\frac{\int dr\,\psi_n^2(r)\,r^4}
     {\int dr\,\psi_n^2(r)\,r^2}
\right]^{1/2},
\ee
in Table~\ref{tab_0++}.

\begin{table}[h!]
    \centering
    \begin{tabular}{|c|c|c|c|c|} \hline
      $E_n$ (GeV) & 1.92 & 2.79 & 3.29 & 4.0 \\
      $r_{\rm r.m.s.}$ ($\mathrm{GeV}^{-1}$) & 2.03 & 5.0 & 6.26 & 6.25 \\ \hline
    \end{tabular}
    \caption{Energies and r.m.s.\ radii of scalar glueballs obtained from the
    Schr\"odinger equation with the parameters defined in the text.}
    \label{tab_0++}
\end{table}

Fixing the model parameters from the masses of the ``normal''
$0^{++}$ states with $n=1,2,3$, which have sizes
$r_{\rm r.m.s.}\sim 5\,\mathrm{GeV}^{-1}\sim 1\,\mathrm{fm}$, we can then
consider other channels. The predicted masses for the higher-$J$ states
with $J=4,6$ turn out to be close to the lattice values. This is expected,
since the large centrifugal barrier increases their spatial extent and
further suppresses sensitivity to short-distance forces. When discussing
these states one should keep in mind that, unlike for mesons, the
confining potential for glueballs is expected to be screened at large
distances, rendering sufficiently large states unstable, formally
corresponding to complex energies. This issue is avoided here by placing
the system inside a spherical cavity: in our calculations we take a
radius $R = 10\,\mathrm{GeV}^{-1}\approx 2\,\mathrm{fm}$ and impose
Dirichlet boundary conditions.

\subsection{The lowest $0^{++}$ scalar glueball}

Here we turn to the lowest $0^{++}$ state, with $n=0$. As discussed in
the Introduction, this state has a long history, with an r.m.s.\ radius
predicted in \cite{Schafer:1994fd} to be
$r_{\rm r.m.s.}(0^{++},n=0)\approx 1\,\mathrm{GeV}^{-1}\approx 0.2\,\mathrm{fm}$,
a result confirmed three decades later in \cite{Abbott:2025irb}. As is
clear from results of our preliminary calculations (Table~\ref{tab_0++}), neither the calculated mass nor the size
of this state agrees with the lattice results. This discrepancy is expected, since those calculation have not yet included the
short-distance attractive spin–spin forces,  perturbative and
instanton-induced. To estimate their combined effect via local interaction
\be
V_{\rm local}(r)=G\,\delta^{3}(r).
\ee
After doing so, we find that reproducing Meyer’s value
$M_{0^{++}}=1.475\,\mathrm{GeV}$ requires
\[
G \approx 38\,\mathrm{GeV}^{-2}.
\]
The corresponding r.m.s.\ radius is reduced, but only to
$r_{\rm r.m.s.}(0^{++},n=0)\approx 1.6\,\mathrm{GeV}^{-1}$, which is still
significantly larger than expected. We therefore conclude that the
lowest scalar glueball is unlikely to be described reliably within the
Schr\"odinger framework, and we will instead treat it below using a
relativistic Bethe–Salpeter approach.

\subsection{Pseudoscalar $0^{-+}$ glueballs}

%The pseudoscalar glueball channel 
%requires separate discussion, since its
%quantum numbers forbid a direct continuation of the leading even-$J$
%adjoint-string-like trajectories. 

In the constituent-gluon picture,
$C=+$ glueballs are composed of two transverse gluons bound by an adjoint
flux tube. Parity and charge conjugation are determined by the orbital
angular momentum $L$ and total gluon spin $S$ according to
\begin{equation}
P = (-1)^{L+1},
\qquad
C = (-1)^{L+S}.
\end{equation}
For $J=0$ and $C=+$, the pseudoscalar quantum numbers $0^{-+}$ require
$L=1$ and $S=1$, corresponding to a ${}^3P_0$ configuration. Thus, unlike
the scalar $0^{++}$ glueball, the $0^{-+}$ state is intrinsically an
orbital excitation. 

(It does not lie on the same trajectory as the
$0^{++}$, $2^{++}$, and higher even-$J$ states. In Regge phenomenology, it correspond to
``odderon" trajectory.)

According to \cite{Meyer:2004gx}, the masses of the two lowest $0^{-+}$
states are
\bea
M_{0^{-+}}^{0}&=&2.250\,(60)(100)\,{\rm GeV},
\nonumber\\
M_{0^{-+}}^{1}&=&3.370\,(150)(150)\,{\rm GeV}.
\eea
Solving the Schr\"odinger equation with $L=1$ and the shifted potential
$V_{\rm conf}+V_0$ (as done for the $0^{++}$ channel), we obtain
\[
M_{0^{-+}}^{0}=2.65\,{\rm GeV},
\qquad
M_{0^{-+}}^{1}=3.13\,{\rm GeV}.
\]
Once again, the excited state is reproduced reasonably well, while the
lowest state shows a noticeable discrepancy.

This discrepancy should originate from short-distance effects, but the
situation in this channel is subtle. The perturbative spin–spin
interaction in the $S_{\rm tot}=1$ channel is attractive, since
$(\vec S_1\!\cdot\!\vec S_2)=-1$, unlike the value
$(\vec S_1\!\cdot\!\vec S_2)=+1/2$ in $L=S=1$ charmonium. In contrast, the
instanton-induced interaction is repulsive (see, for example,
Fig.~\ref{fig_corr}). Moreover, for nonzero $L$ one has $\psi(0)=0$, so
the perturbative contribution depends sensitively on the smearing of the
delta function. For these reasons, we leave this issue unresolved.

\subsection{WKB baseline for the $0^{-+}$ channel}

The unshifted WKB mass for the $0^{-+}$ glueball is obtained by identifying
$J=L=1$ in Eq.~\eqref{eq:WKBspec}. For the radial quantum number $n=0$, this 
amounts to
\begin{equation}
M^{\rm WKB}_{0^{-+}}
=
2m_g
+
\left[
\frac{3\pi}{2}
\left(
\frac{1}{2}+\frac{3}{4}
\right)
\right]^{2/3}
\sigma_8^{2/3} m_g^{-1/3}.
\label{eq:WKB_0minus}
\end{equation}
At $\eta=1$ this value is numerically comparable to the unshifted $J=2$
tensor baseline. The corresponding turning point is
$r_+=E/\sigma_8$, with $E=M^{\rm WKB}-2m_g$.

The resulting WKB estimates for the lowest pseudoscalar states are listed
in Table~\ref{tab:pseudoscalar_WKB}. The quoted radii provide
semiclassical measures of the spatial extent and should be interpreted in
the same spirit as those in Table~\ref{tab_corr}.

\begin{table}[h!]
\centering
\begin{tabular}{cccccc}
\hline\hline
$n$ & $L$ & $J^{PC}$ & $M^{\rm WKB}_{nL}$ (GeV) & $r_+$ (fm) & $r^{\rm WKB}_{\rm rms}$ (fm) \\
\hline
0 & 1 & $0^{-+}$ & 3.33 & 1.27 & 0.93 \\
1 & 1 & $0^{-+}$ & 4.56 & 1.86 & 1.36 \\
2 & 1 & $0^{-+}$ & 5.59 & 2.36 & 1.72 \\
\hline\hline
\end{tabular}
\caption{Semiclassical WKB baseline for pseudoscalar $0^{-+}$ glueballs at
$\eta=1$. The $0^{-+}$ channel necessarily carries $L=1$ and therefore
starts at a higher mass than the scalar trajectory. The turning point
$r_+$ and r.m.s.\ radius provide semiclassical measures of the spatial
extent of the states.}
\label{tab:pseudoscalar_WKB}
\end{table}

\subsection{Quantum numbers of other glueballs }
\label{sec:vector}

Let us remind 
standard classification of the two-gluon quantum states.
Consider two  spin-1  effective constituent gluons (bosons) in a color singlet $8\otimes 8$ 
representation. It is symmetric under exchange, hence the remaining part of the wave function
(spin$\times$space)  must also be symmetric.
Writing the total wavefunction under particle exchange as
\begin{equation}
\mathcal{P}_{12}\Psi = (+1)\Psi,
\end{equation}
the spatial part contributes $(-1)^L$, while the spin part is symmetric for
$S=0,2$ and antisymmetric for $S=1$.  Therefore symmetry requires
\begin{equation}
(-1)^L = +1 \quad \text{for } S=0,2,
\qquad
(-1)^L = -1 
%\quad \text{for } S=1.
\label{eq:sym_constraint}
\end{equation}
for $S=1$.
Now we consider whether a $J=1$ state can be formed from two such bosons.
The possible angular couplings to $J=1$ are
\begin{equation}
(L,S)=(0,1),\ (1,0),\ (1,2),\ (2,1),\dots
\end{equation}
but each is excluded by \eqref{eq:sym_constraint}:
$(0,1)$ has $L$ even but requires $S=1$ (antisymmetric), hence forbidden;
$(1,0)$ and $(1,2)$ have $L$ odd but require $S=0$ or $S=2$ (symmetric), hence forbidden;
$(2,1)$ has $L$ even but requires $S=1$, hence forbidden.
Thus, in a symmetric color-singlet two-gluon configuration,
 no $J=1$ state exists in the two-gluon sector.
(This  version of the Landau-Yang selection rule carries to massive gluons as well.)

Charge conjugation gives  additional constraints.  Two identical gluons in a color
singlet have
\begin{equation}
C(gg)=+,
\label{eq:Cplus}
\end{equation}
so true vectors $1^{-}$ (with $C=-$) are not accessible in the leading two-gluon
Fock sector at all.  The pseudovector $1^{+-}$ has $C=+$, but is excluded by
the preceding argument.  Therefore the lowest vector and pseudovector glueballs are expected
to be dominated by multi-gluon (especially three-gluon) structure rather than a tightly
bound two-gluon core.

On top of these kinematical constraints, there are effects of dynamical origin related to  specific self-dual structure of the instanton fields. Although
classical fields are strong, all
components of the stress tensor 
are zero at all points. As a result, ILM predicted \cite{Schafer:1994fd} suppressed
nonperturbative effects in the tensor $2^{++}$ channel.

This suppression needs to be modified in a version of ILM including instanton-antiinstanton
molecules. They are not 
strictly self-dual, and contain
rotating Pointing vectors, which can in principle affect the tensor channel.

%The vector channels are sensitive  to the forces implied by
%the ILM, as well as the constraint from
%Bose symmetry.  In the ILM, the strongest attractive contributions arise in channels
%that couple efficiently to (anti)self-dual field strengths and their lowest-dimension
%composites, which favors the scalar (and to some extent the tensor and pseudoscalar).
%By contrast, vector channels are not favored: the instanton-induced scalar-core attraction
%is absent or weak, and the dominant short-distance contributions are reduced by
%symmetry constraints and centrifugal suppression. 
The ILM
suggests that the lightest vector channels are weakly bound, or mostly unbound as
two-constituent $gg$ states.
In our
Hamiltonian, the short-distance attractions behave parametrically as
\begin{equation}
\langle V_C\rangle \sim -3\alpha_s\Big\langle\frac{1}{r}\Big\rangle,
\qquad
\langle V_{\rm inst}\rangle \sim -G\,\Big\langle e^{-r^2/\rho^2}\Big\rangle,
\end{equation}
Both require significant probability density at small distances $r\lesssim\rho$.
Yet any state forced into $L\ge1$ is suppressed near the origin by the centrifugal barrier,
so $\langle 1/r\rangle$ and $\langle e^{-r^2/\rho^2}\rangle$ are reduced.   Spin-dependent terms are likewise reduced because the regulated contact
overlap decreases rapidly with increasing $L$.  Consequently, even if confinement
supports high-lying excitations, the ILM does not generically provide 
additional attraction needed for a low, compact vector glueball.

This qualitative picture is consistent with quenched SU(3) lattice spectroscopy,
which finds the lightest pseudovector $1^{+-}$ substantially heavier than the
tensor and scalar, and the lightest $1^{-}$ heavier still~\cite{Morningstar:2025glueballreview}.  In this sense,
the vector channels provide evidence that these glueballs 
do not experience
 a universal two-body short-range attraction, existing in selected $C=+$ channels (notably the scalar) together with
confinement and multi-gluon dynamics.

\section{Relativistic treatment of short-distance dynamics}
\label{sec:relativistic}
\subsection{Emergent 4-gluon interaction}

%E there are noe fermions, next
%para not needed

%At fixed pseudoparticle moduli, the low-mode fermion determinant generates the one-(anti)instanton induced vertices (for each instanton $I$ and anti-instanton $A$) dressed with perturbative gluons of the form~\cite{Liu:2024glue}
%\begin{widetext}
%\begin{align}
%\Theta_{I}
%&=\prod_{f}\Bigg[
%\frac{m_{f}}{4\pi^{2}\rho^{2}}
%+i\,\psi_f^{\dagger}(x)\,U_{I}\,
%\frac{1}{2}\Big(1+\frac{1}{4}\tau^{a}\,\bar\eta^{a}_{\mu\nu}\sigma_{\mu\nu}\Big)\,
%U_{I}^{\dagger}\,\frac{1-\gamma_{5}}{2}\,\psi_f(x)
%\Bigg]\,
%\exp\!\Big[-%\kappa\,\rho^{2}\,R^{ab}(U_{I})\,\bar\eta^{b}_{\mu\nu}\,G^{a}_{\mu\nu}(x)\Big],\nonumber\\
%\Theta_{A}
%&=\prod_{f}\Bigg[
%\frac{m_{f}}{4\pi^{2}\rho^{2}}
%+i\,\psi_f^{\dagger}(x)\,U_{A}\,
%\frac{1}{2}\Big(1+\frac{1}%{4}\tau^{a}\,\eta^{a}_{\mu\nu}\sigma_{\mu\nu}\Big)\,
%U_{A}^{\dagger}\,\frac{1+\gamma_{5}}{2}\,\psi_f(x)
%\Bigg]\,
%\exp\!\Big[-\kappa\,\rho^{2}\,R^{ab}(U_{A})\,\eta^{b}_{\mu\nu}\,G^{a}_{\mu\nu}(x)\Big],
%\label{eq:Theta45_rewrite}
%\end{align}
%\end{widetext}
%in the local (zero size) approximation, with $\kappa\equiv 2\pi^{2}/g$, and 
%$$R^{ab}(U)=\frac{1}%{2}\Tr(\tau^{a}U\tau^{b}U^{\dagger})$$.

We begin by sketching derivation
of effective multigluon vertices induced by instantons.
It starts by
an exponential of the (LSZ-reduced) gluonic ``tail emission'' from the pseudoparticle, written compactly as a coupling to the field strength $G^a_{\mu\nu}$.

%\subsection{Expansion}

For the present purpose we isolate the purely gluonic source factor in \eqref{eq:Theta45_rewrite}, and restaure the
finite size form factor
\begin{widetext}
\begin{eqnarray}
\mathcal{E}_I[G]=
\exp\!\left[
-\kappa\rho^2
\int\frac{d^4q}{(2\pi)^4}e^{iq\cdot x}
\beta_{2g}(\rho|q|)
R^{ab}(U)\bar\eta^b_{\mu\nu}G^a_{\mu\nu}(q)
\right],
\nonumber\\
\mathcal{E}_A[G]=
\exp\!\left[
-\kappa\rho^2
\int\frac{d^4q}{(2\pi)^4}e^{iq\cdot x}
\beta_{2g}(\rho|q|)
R^{ab}(U)\eta^b_{\mu\nu}G^a_{\mu\nu}(q)
\right].
\nonumber\\
\label{eq:E_def}
\end{eqnarray}
\end{widetext}
%with $\kappa\equiv \frac{2\pi^{2}}{g}.$
For a single (anti)-instanton of size $\rho$ centered at the origin, the field strength is
\begin{eqnarray}
    {\mathbb G}_{\mu\nu}^a(x)=\frac 4g\,\bar{\eta}^a_{\mu\nu}\frac{\rho^2}{(x^2+\rho^2)^2}
\end{eqnarray}
and the field-strength Fourier transform or form factor, is
\begin{equation}
{\mathbb G}_{\mu\nu}^a(q)=\frac{4\pi^2\rho^2}g\bar\eta_{\mu\nu}^a\,\beta_{2g}(t)
\equiv \frac{4\pi^2\rho^2}g\bar\eta_{\mu\nu}^a\,\bigg(\frac{t^2}{2}K_2(t)\bigg)\,.
\end{equation}
It is gauge invariant, satisfies $\beta_{2g}(0)=1$ and decays exponentially for $t\gg1$. An analogous expression holds for anti-instantons with $\bar\eta\to\eta$. The extra $\frac 12$ that appears in the exponential arises from the  coupling to the background field
\begin{eqnarray}
    &&\frac 14\int d^4x\,2\,{\mathbb G}_{\mu\nu}^a(x)G_{\mu\nu}^a(x)\nonumber\\
    &&\rightarrow \frac 12
    G^a_{\mu\nu}(q)\int d^4x\,{\mathbb G}^a_{\mu\nu}(x)e^{-iq\cdot x}
\end{eqnarray}
where the rightmost equation follows by LSZ reduction of the perturbative gluon. Note that this reconstruction of the exponent is fully gauge invariant.

The instanton-induced four-gluon operator arises from the fourth-order term in the expansion of the tail-emission functional. Writing
\begin{equation}
X_I[G]=\kappa\rho^2\int\frac{d^4q}{(2\pi)^4}e^{iq\cdot x}
\beta_{2g}(\rho|q|)
R^{ab}(U)\bar\eta^b_{\mu\nu}G^a_{\mu\nu}(q),
\end{equation}
and similarly for $X_A[G]$, 
the quartic contribution is proportional to
\begin{equation}
n\left\langle \frac{X_I^4}{4!}+\frac{X_A^4}{4!}\right\rangle_U,
\end{equation}
where $n$ is the instanton density and $\langle\cdots\rangle_U$ denotes averaging over color orientations. 
%with
%\begin{equation}
%\frac{X^4}{4!}
%=
%\frac{(\kappa\rho^2)^4}{4!}
%\Big[\prod_{i=1}^4 R_{a_i b_i}(U)\Big]
%\Big[\prod_{i=1}^4 \bar\eta^{\,b_i}_{\mu_i\nu_i}\,
%G^{a_i}_{\mu_i\nu_i}(x)\Big].
%\label{eq:X4_raw_new}
%\end{equation}
Averaging over color orientation with the Haar measure projects onto singlets. For the adjoint dimension $d_A=N_c^2-1$, the fourth group average yields the standard contraction structure
\begin{widetext}
\begin{equation}
\Big\langle R_{a_1 b_1}R_{a_2 b_2}R_{a_3 b_3}R_{a_4 b_4}\Big\rangle_U
\;\Rightarrow\;
\frac{3}{d_A(d_A+1)}\,
\Big(\delta_{a_1 a_2}\delta_{b_1 b_2}\Big)
\Big(\delta_{a_3 a_4}\delta_{b_3 b_4}\Big)
+\text{perm.},
\label{eq:R4_avg_tensor_new}
\end{equation}
\end{widetext}
To proceed, we recall the identity
\begin{equation}
\bar\eta^{\,b}_{\mu\nu}\bar\eta^{\,b}_{\rho\sigma}
=
\delta_{\mu\rho}\delta_{\nu\sigma}
-
\delta_{\mu\sigma}\delta_{\nu\rho}
-
\epsilon_{\mu\nu\rho\sigma},
\label{eq:eta_id_new}
\end{equation}
which yields to the contractions
\begin{equation}
\bar\eta^{\,b}_{\mu\nu}\bar\eta^{\,b}_{\rho\sigma}
G^{a}_{\mu\nu}G^{a}_{\rho\sigma}
=
2\,G^{a}_{\mu\nu}G^{a}_{\mu\nu}
-
2\,G^{a}_{\mu\nu}\tilde G^{a}_{\mu\nu}.
\label{eq:G2_decomp_new}
\end{equation}
Projecting onto the parity-even scalar channel retains only the $(G^a_{\mu\nu}G^a_{\mu\nu})^2$ part, leading the induced and non-local scalar four gluon interaction
\begin{widetext}
\begin{equation}
\Delta\mathcal{L}_{4g}
=
n\,
\frac{\kappa^4\rho^8}{2\,d_A(d_A+1)}
\int\prod_{i=1}^4\frac{d^4q_i}{(2\pi)^4}\,
\beta_{2g}(\rho|q_i|)
(2\pi)^4\delta^{(4)}\!\left(\sum_i q_i\right)
\big[G^a_{\mu\nu}(q_1)G^a_{\mu\nu}(q_2)\big]
\big[G^c_{\rho\sigma}(q_3)G^c_{\rho\sigma}(q_4)\big]
+\text{perm.}
\label{eq:DeltaL4g_final_newx}
\end{equation}
\end{widetext}
Each external gluon momentum carries its own form factor $\beta_{2g}(\rho|q_i|)$. Note the leg-by-leg factorization which will be key for the bound state equation to follow. In the local approximation (zero size) the 4-gluon vertex is
\begin{equation}
\Delta{\cal L}^{(0^{++})}_{4g}(x)
=
n\,
\frac{\kappa^{4}\rho^{8}}{2\,d_A(d_A+1)}\,
\Big(G^{a}_{\mu\nu}(x)G^{a}_{\mu\nu}(x)\Big)^2.
\label{eq:DeltaL4g_final_new}
\end{equation}
Eq.~\eqref{eq:DeltaL4g_final_newx} is the microscopic one-(anti)instanton contact vertex in the $0^{++}$ channel that we will now reduce to an instantaneous two-gluon potential.

\subsection{Emergent 2-gluon $0^{++}$-potential}
To extract the gluon two-body  potential in the scalar $0^{++}$ channel, we will use on-shell
in-out gluonic external states and the non-local operator \eqref{eq:DeltaL4g_final_newx}. This procedure parallels the one discussed by us~\cite{Shuryak:2021fsu} for the constituent quark pair interactions. 

For the  on-shell one-gluon states in the CM frame, we use the normalization
\begin{equation}
\langle \vec k,\lambda,a | \vec k',\lambda',a'\rangle
=
(2\pi)^3\,2\omega\,
\delta^{(3)}(\vec k-\vec k')\,
\delta_{\lambda\lambda'}\,\delta_{aa'}.
\label{eq:gluon_norm_new}
\end{equation}
For a physical transverse polarization vector $\epsilon^\mu(k,\lambda)$, the field-strength matrix element is
\begin{equation}
\langle 0|\,G^a_{\mu\nu}(0)\,|g^b(k,\lambda)\rangle
=
i\,\delta^{ab}\,
\big(k_\mu\epsilon_\nu(k,\lambda)-k_\nu\epsilon_\mu(k,\lambda)\big),
\label{eq:G_ext_new}
\end{equation}
and similarly for outgoing legs with $\epsilon\to\epsilon^{*}$. Define the gauge-invariant two-gluon contraction induced by $G^a_{\mu\nu}G^a_{\mu\nu}$,
\begin{widetext}
\begin{equation}
\mathcal{F}(i,j)\equiv
G^a_{\mu\nu}(i)\,G^a_{\mu\nu}(j)
\;\Rightarrow\;
-\,2\,\delta^{a_i a_j}\Big[(k_i\!\cdot\!k_j)(\epsilon_i\!\cdot\!\epsilon_j)
-(k_i\!\cdot\!\epsilon_j)(k_j\!\cdot\!\epsilon_i)\Big].
\label{eq:Fij_new}
\end{equation}
At Born level, \eqref{eq:DeltaL4g_final_new} yields
\begin{align}
i\mathcal{M}^{(4)}(1,2\to3,4)
=
i\,n\,
\frac{\kappa^4\rho^8}{2\,d_A(d_A+1)}\prod_{i=1}^4\beta_{2g}(\rho|k_i|)
%\Big[
%\beta_{2g}(\rho|k_1|)
%\beta_{2g}(\rho|k_2|)
%\beta_{2g}(\rho|k_3|)
%\beta_{2g}(\rho|k_4|)
%\Big]
%\nonumber\\
%&\quad\times
\Big[
\mathcal{F}(1,3)\mathcal{F}(2,4)
+
\mathcal{F}(1,4)\mathcal{F}(2,3)
\Big]\,.
\label{eq:M4_corrected}
\end{align}
\end{widetext}
Projecting onto the two-gluon color singlet state $|1\rangle_{\rm color}=\delta^{ab}|ab\rangle/\sqrt{d_A}$ sets the color factors in \eqref{eq:M4_corrected} to unity for both exchange structures.

The scalar $0^{++}$ projection is implemented by taking the $J_z=0$ helicity combination of two transverse gluons,
\begin{equation}
|0^{++}\rangle_{\rm pol}
=
\frac{1}{\sqrt{2}}
\Big(|\lambda_1{=}{+},\lambda_2{=}{-}\rangle
+
|\lambda_1{=}{-},\lambda_2{=}{+}\rangle\Big),
\label{eq:scalar_pol_state_new}
\end{equation}
and similarly for the outgoing state. In CM kinematics with scattering angle $\theta$ between $\vec k_1$ and $\vec k_3$, the detail contractions of the transverse helicities for massive on-shell gluons yield
\begin{widetext}
\begin{equation}
\langle 0^{++}|\,
\Big[\mathcal{F}(1,3)\mathcal{F}(2,4)+\mathcal{F}(1,4)\mathcal{F}(2,3)\Big]
\,|0^{++}\rangle
=2\,(\omega^2+k^2)^2\,\Big(1+\cos^{2}\theta\Big),
\label{eq:pol_factor_theta_new}
\end{equation}
\end{widetext}
as detailed in Appendix~\ref{sec:2++}.

Using the Legendre polynomials,
\begin{equation}
1+\cos^2\theta
=
\frac{4}{3}\,P_0(\cos\theta)+\frac{2}{3}\,P_2(\cos\theta),
\label{eq:L_Pdecomp}
\end{equation}
the decomposition apparently shows both $J=0,2$. 
If we regard \eqref{eq:L_ang} as a helicity-0 object and performs a
spinless partial-wave projection,
\begin{equation}
\mathcal M^{(4)}_{J}
=
\frac{2J+1}{2}\int_{-1}^{1}dx\,
P_J(x)\,\mathcal M^{(4)}(x),
\qquad x=\cos\theta,
\label{eq:L_spinlessPW}
\end{equation}
then \eqref{eq:L_Pdecomp} implies the  weights
\begin{equation}
\mathcal K_{0^{++}}=\frac{4}{3},
\qquad
\mathcal K_{2^{++}}=\frac{2}{3},
%\qquad
%\frac{\mathcal K_{2^{++}}}{\mathcal K_{0^{++}}}=\frac{1}{2}.
\label{eq:K00_new}
%\label{eq:L_half}
\end{equation}
hence
$$\frac{\mathcal K_{2^{++}}}{\mathcal K_{0^{++}}}=\frac{1}{2}.$$
However, this ratio  reflects only the relative size of the $P_2$ and $P_0$ components in
\eqref{eq:L_Pdecomp}, with the latter  essentially an S-channel amplitude. While  $\mathcal K_{0^{++}}$ is the proper weight in the iteration  of the $0^{++}$, $\mathcal K_{2^{++}}$ does not carry the correct
weight in the $2^{++}$ as detailed in Appendix~\ref{sec:2++}. Although the emergent instanton  vertex contains a $P_2$ component with relative weight $1/2$ in the helicity-0 decomposition, the \emph{physical} $2^{++}$ interaction built from transverse gluons follows from   helicity projection as we detail 
in Appendix~\ref{sec:2++}.

With this in mind, the in-out legs form factors in the S-channel can be restaured as follows. In the glueball rest frame, the incoming legs carry relative momentum $\bm p$ and the outgoing legs carry $\bm k$. The four form factors combine as
\begin{eqnarray}
\prod_{i=1}^4\beta_{2g}(\rho|k_i|)
%&&\beta_{2g}(\rho|k_1|)\beta_{2g}(\rho|k_2|)\beta_{2g}(\rho|k_3|)\beta_{2g}(\rho|k_4|)
%\nonumber\\
%&&=
=\beta_{2g}^2(\rho p)\beta_{2g}^2(\rho k).
\end{eqnarray}
Since the exchange is taking place inside an instanton (anti-instanton) this is appropriately described by an {\it instantaneous} scalar potential, hence  
\begin{equation}
V(\bm p,\bm k)
=
-\,V_0\,
\beta_{2g}^2(\rho p)\,
\beta_{2g}^2(\rho k).
\end{equation}
which is separable.
The strength $V_0$ is obtained by combining the quartic coefficient, the scalar projection factor, and the normalization required to convert an invariant amplitude into a three-dimensional potential. The latter introduces reduced-state factors $(2\omega)^{-1/2}$ per external leg. Matching these factors at the dominant instanton scale $p,k\sim\rho^{-1}$ yields
\begin{equation}
\omega_\rho=\sqrt{\rho^{-2}+m_g^2},
\end{equation}
and produces a factor $(2\omega_\rho)^{-2}=1/(4\omega_\rho^2)$ in the effective coupling. Collecting all contributions gives
\bea
V_0
=
n\,
\frac{\kappa^4}{2\,d_A(d_A+1)}
\;\frac{16{\mathcal K}_{0^{++}}}{4\omega_\rho^2}\,
\rho^4.\\\nonumber
\eea

\subsection{Reduced Bethe-Salpeter equation}

Let $\Gamma(p;P)$ be the amputated Bethe-Salpeter (BS) vertex for two equal-mass constituent gluons of mass $m_g$, with total four-momentum $P$ and relative four-momentum $p$. In the ladder approximation with an instantaneous kernel $V(\bm p,\bm k)$, the reduced BS (Salpeter) equation for the scalar $0^{++}$ glueball is
\begin{widetext}
\begin{equation}
\Gamma(p;P)
=
\int\frac{d^4k}{(2\pi)^4}\;
%K(p,k;P)\;
V(\bm p,\bm k)\;
D\!\left(k+\frac{P}{2}\right)\;
D\!\left(-k+\frac{P}{2}\right)\;
\Gamma(k;P),
\label{eq:BS_general}
\end{equation}
\end{widetext}
where $-iD=1/(q^2-m_g^2+i0)$ is the constituent-gluon propagator. .

In Eq.~\eqref{eq:BS_general} the interaction kernel is constructed using on-shell transverse gluons.
This reflects the instantaneous (Salpeter) reduction to follow: after integrating
over the relative energy, the dominant contributions arise from the poles of the gluon
propagators, and the bound state below threshold is governed by physical transverse
degrees of freedom. Off-shell and gauge components are usually implicitly encoded in the effective kernel of the Salpeter reduction, and do not affect the mass eigenvalue. In our case the instanton induced pair interaction is
gauge invariant by construction.

%\begin{equation}
%D(q)=\frac{i}{q^2-m_g^2+i0}.
%\end{equation}
With this in mind and in the glueball rest frame
\begin{equation}
P^\mu=(M,\bm 0),\qquad p^\mu=(p_0,\bm p),\qquad k^\mu=(k_0,\bm k),
\end{equation}
we define the equal-time Salpeter amplitude
\begin{equation}
\phi(\bm p)\equiv \int\frac{dp_0}{2\pi}\;
D\!\left(p+\frac{P}{2}\right)\;
D\!\left(-p+\frac{P}{2}\right)\;
\Gamma(p;P).
\label{eq:Phi_def}
\end{equation}
Since $V(\bm p,\bm k)$ does not depend on $p_0$ or $k_0$, we can multiply both sides 
of \eqref{eq:BS_general}  by the product of propagators and integrate over $p_0$ as in \eqref{eq:Phi_def}. This yields
\begin{widetext}
\begin{equation}
\phi(\bm p)
=
\int\frac{d^3k}{(2\pi)^3}\;
V(\bm p,\bm k)\;
\left[
\int\frac{dp_0}{2\pi}\,
D\!\left(p+\frac{P}{2}\right)\;
D\!\left(-p+\frac{P}{2}\right)
\right]\;
\phi(\bm k)
\label{eq:Phi_intermediate}
\end{equation}
%\end{widetext}
The remaining $p_0$ integral is elementary,
%\begin{widetext}
\begin{equation}
\int\frac{dp_0}{2\pi}\,D\!\left(p+\frac{P}{2}\right)D\!\left(-p+\frac{P}{2}\right)
=\int\frac{dp_0}{2\pi}\,
\prod_{s=\pm}\frac{i}{(p_0+sM/2)^2-\omega_p^2+i0}
%\frac{i}{(p_0-M/2)^2-\omega_p^2+i0}
=\frac{1}{2\omega_p}\;
\frac{1}{M^2-4\omega_p^2+i0},
%\qquad \omega_p\equiv \sqrt{p^2+m_g^2},
\end{equation}
\end{widetext}
 by contour integration, hence 
the reduced Bethe-Salpeter integral equation
\begin{equation}
\phi(\bm p)
=
\frac{1}{2\omega_p}\;
\frac{1}{M^2-4\omega_p^2}\;
\int\frac{d^3k}{(2\pi)^3}\;
V(\bm p,\bm k)\;
\phi(\bm k),
\label{eq:Salpeter_3D}
\end{equation}

\subsection{The scalar glueball $0^{++}$ mass equation}
Substituting the separable kernel yields the rank-1 integral equation for the scalar glueball $0^{++}$ wavefunction in the CM frame
\begin{equation}
\phi(\bm p)=-
\,V_0
\frac{\beta_{2g}^2(\rho p)}{2\omega_p
(M^2-4\omega_p^2+i0)}
\int\frac{d^3k}{(2\pi)^3}
\beta_{2g}^2(\rho k)\phi(\bm k).
\end{equation}
Eliminating the normalization constant,  leads to the  root equation
\begin{equation}
1=
V_0
\,{\bf PP}\,\int_0^\infty\frac{p^2\,dp}{(2\pi)^2}
\frac{\beta_{2g}^4(\rho p)}{2\omega_p(4\omega_p^2-M^2)}.
\end{equation}
with the Principal Part (${\bf PP}$) retained for a bound state. 
Introducing $x=\rho p$, $\mu=m_g\rho$, and $\mathcal{M}=M\rho$ yields the dimensionless form
\begin{equation}
1=
\lambda_M\,{\bf PP}\,
\int_0^\infty dx
\frac{x^2}{\sqrt{x^2+\mu^2}}
\frac{\left[\frac{x^2}{2}K_2(x)\right]^4}{4(x^2+\mu^2)-\mathcal{M}^2},
%\qquad
%\lambda_E=\frac{V_0\rho^2}{(2\pi)^2}.
\label{eq:root_mass}
\end{equation}
with
$$\lambda_M=\frac{V_0\rho^2}{(2\pi)^2}.$$
Eq.~\eqref{eq:root_mass} is the explicit rank-1 root mass equation for the $0^{++}$ glueball rescaled mass.

Finally, note that the bound-state mass is fixed by the pole condition of the two-body Green function,
$1-V_0\Pi(P^2)=0$, which is analytic below threshold. Hence, the Minkowski equation
evaluated at $P^2=M^2$ and the Euclidean equation evaluated at $P_E^2=-M^2$, yield
the same root mass \eqref{eq:root_mass} by analytical continuation, which in Euclidean signature reads
\begin{equation}
1=
\lambda_M
\int_0^\infty dx
\frac{x^2}{\sqrt{x^2+\mu^2}}
\frac{\left[\frac{x^2}{2}K_2(x)\right]^4}{4(x^2+\mu^2)+\mathcal{M}^2},
%\qquad
%\lambda_E=\frac{V_0\rho^2}{(2\pi)^2}.
\label{eq:root_massE}
\end{equation}

\subsection{The scalar glueball  wavefunction}

In contrast, the reduced momentum-space wavefunctions in Minkowski and Euclidean space are different, yet related to each other. Indeed, 
in the equal-time (Minkowski) reduction the wavefunction in momentum space is explicitly  given by
\begin{equation}
\phi_M(\mathbf p)={\cal N}_M
\frac{\beta_{2g}^2(\rho p)}
{2\omega_p\,(M^2-4\omega_p^2+i0)} ,
\label{eq:wf_momentum_CM}
\end{equation}
In contrast, the Euclidean reduced Bethe-Salpeter amplitude is a 4-dimensional function of the form 
\begin{equation}
\chi_E(p_4,\mathbf p)
\;\sim\;
\frac{\beta_{2g}^2(\rho p)}
{\big[(p_4+\tfrac{P_4}{2})^2+\omega_p^2\big]
 \big[(p_4-\tfrac{P_4}{2})^2+\omega_p^2\big]},
%\qquad
%\omega_p=\sqrt{p^2+m_g^2},
\end{equation}
originating from the product of the two Euclidean propagators in the iterating kernel,
with no additional factor $1/(2\omega_p)$. The equal-time (Salpeter) wavefunction follows by slicing over the relative energy,
\begin{equation}
\phi_M(\mathbf p)
=
\int_{-\infty}^{\infty}\frac{dp_4}{2\pi}\,
\chi_E(p_4,\mathbf p)
\;\propto\;
\frac{\beta_{2g}^2(\rho p)}
{2\omega_p\,(M^2-4\omega_p^2+i0)} .
\end{equation}
The extra factor of $1/(2\omega_p)$ arises from the relative-energy integration, and is
absent in the 4-dimensional Euclidean amplitude.

\subsection{Numerical results}
\begin{figure}
    \centering
    \includegraphics[width=0.85\linewidth]{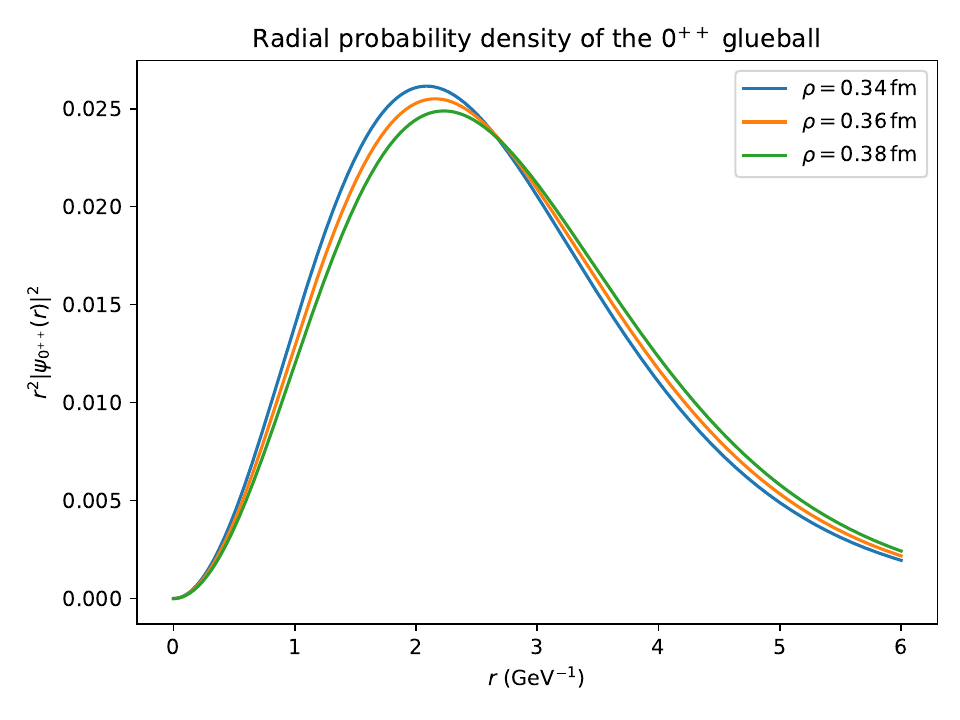}
    \caption{$0^{++}$ Glueball radial probability  $r^2|\psi_{0^{++}}(r)|^2$ versus $r(\rm GeV^{-1})$ from the reduced Bethe-Salpeter equation, for different (anti)instanton size $\rho=0.34, 0.36, 0.38$ fm.}
\label{fig:glueball_radial}
\end{figure}

For sufficiently strong (anti)instanton coupling $\lambda_M$ a bound state below 2-constitutive gluon treshold can form.
Since $\lambda_M\sim\rho^{\,p+2}$ with $p+2\gtrsim6$, this coupling is very sensitive to the (anti)instanton size $\rho$, e.g.
\begin{equation}
\left(\frac{\rho}{0.32~\mathrm{fm}}\right)^6
\sim 10\text{-}15
%\qquad\Rightarrow\qquad
%\rho\simeq0.36\text{-}0.38~\mathrm{fm}.
\end{equation}
for $\rho\simeq0.36\text{-}0.38~\mathrm{fm}.$
With this in mind, for a dense ILM, with
$\eta=6-7$ and
%We now proceed to solve (\ref{eq:root_mass}) using the parameters in the range
\begin{eqnarray}
%&&\rho = 0.30\text{-}0.33~\mathrm{fm},\qquad 
%\eta = 6\text{-}7,\nonumber\\
&&m_g(\eta)=m_{g0}\sqrt{\eta},\quad m_{g0}=0.36~\mathrm{GeV},\nonumber\\
&&\alpha=\frac{g^2}{4\pi}=0.3\text{-}0.5,
\end{eqnarray}
 the scalar $0^{++}$ glueball mass is
\begin{equation}
M_{0^{++}} \;\simeq\; 1.4\text{-}1.5~\mathrm{GeV},
\end{equation}
well  below the two-gluon threshold
$2m_g(\eta)=1.8$-$1.9~\mathrm{GeV}$.
The $0^{++}$ rms radius follows from the exact momentum-space eigenstate
\eqref{eq:wf_momentum_CM}, using
\begin{equation}
\langle r^2\rangle
=
\int\frac{d^3p}{(2\pi)^3}\,
\phi_M^*(\mathbf p)
\left(-\nabla_{\mathbf p}^2\right)
\phi_M(\mathbf p).
\end{equation}
with the result
\begin{equation}
\langle r^2\rangle^{1/2}_{0^{++}}
\simeq
0.25\text{-}0.32~\mathrm{fm},
\end{equation}
This shows a low-lying and compact scalar glueball, dominated by instanton-scale dynamics, as illusttrated by the radial probability distribution shown in Fig.~\ref{fig:glueball_radial}

\section{Glueball spectrum}
\label{subsec:spectrum_PC}

The constituent two-gluon Hamiltonian organizes the glueball spectrum naturally
into radial and orbital families, in close analogy with quarkonium spectroscopy.
This separation is particularly transparent when the instanton density parameter
is fixed at $\eta=1$ or in the ILM, where the scalar and tensor ground states are fitted to their lattice values.

Radial excitations correspond to states with fixed total spin and parity but
increasing number of nodes in the relative wavefunction.
In the scalar channel, the ground state $0^{++}_0$ is strongly shifted downward
by Coulomb and instanton-induced attraction and is compact, with a size of order
the instanton radius.
The first radial excitation $0^{++}_1$ is significantly less affected by
short-distance dynamics because its wavefunction extends to larger radii;
its mass therefore lies closer to the confinement-dominated WKB prediction.
This pattern reflects the rapid decoupling of instanton physics with increasing
radial quantum number.

Orbital excitations instead form Regge-like towers at fixed radial quantum number.
The tensor family $2^{++},4^{++},6^{++},\ldots$ provides the clearest example.
Here the centrifugal barrier suppresses short-distance overlap already at the
ground state, and ${}^5S_2$-${}^5D_2$ mixing further redistributes probability
toward larger radii.
As a result, the tensor ground state is only moderately shifted from the WKB
baseline, and higher-$J$ states follow an approximately linear Regge trajectory
with slope fixed by the adjoint string tension.

\begin{figure*}[t]
\centering
\includegraphics[width=\textwidth]{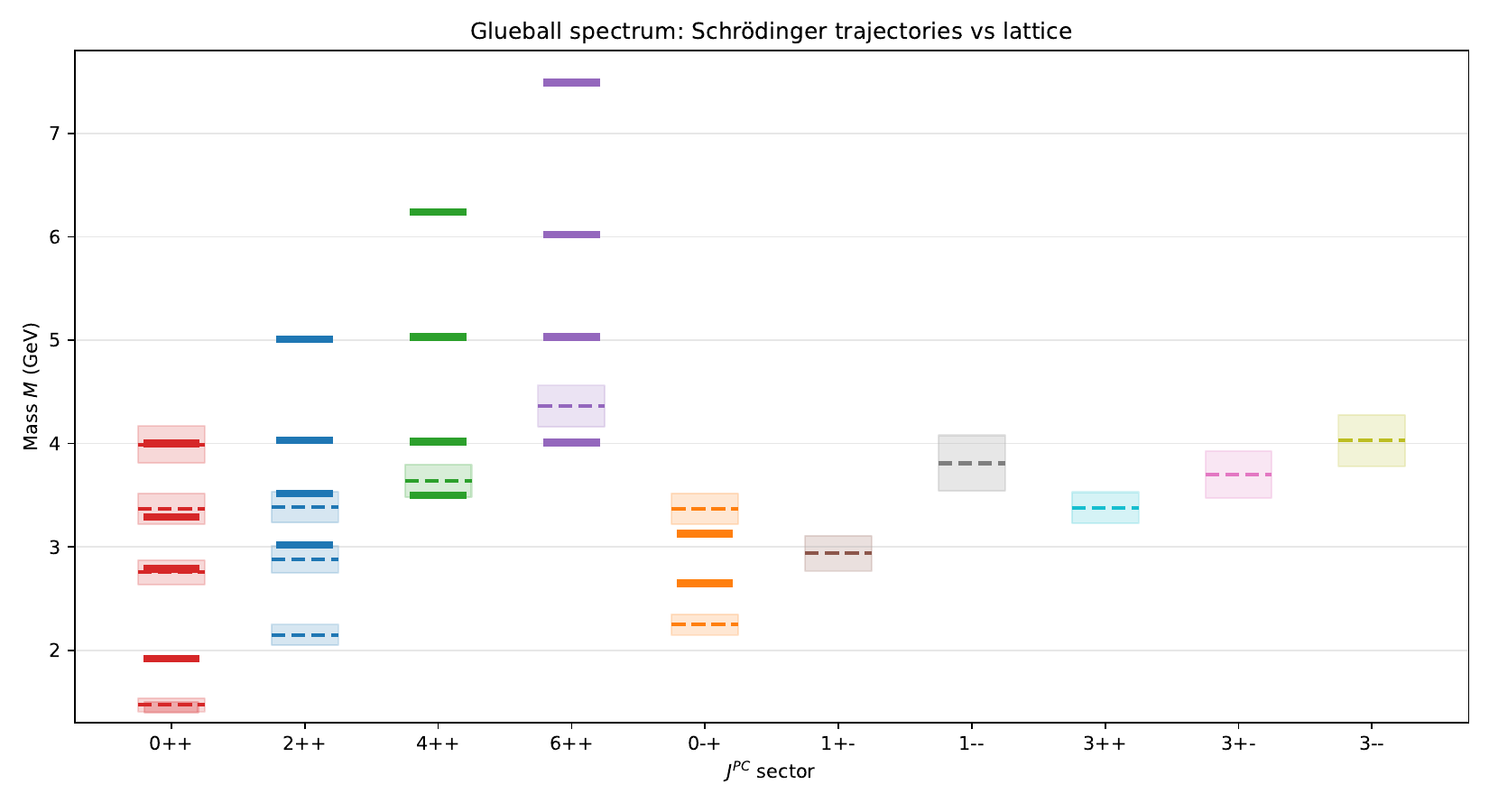}
\caption{
Glueball mass spectrum organized by $J^{PC}$ sectors, as listed in
Table~\ref{tab:glueball_twocol_JPC}. Solid markers (this work) and
dashed (mean) and spread (uncertainty) markers from quenched SU(3) lattice~\cite{Meyer:2004gx,Morningstar:1999rf}.
}
\label{fig:glueball_spectrum_PC}
\end{figure*}

Figure~\ref{fig:glueball_spectrum_PC} presents the glueball spectrum organized by
parity and charge conjugation, following the standard lattice-spectroscopy layout.
The comparison highlights the selective role of instanton-induced dynamics.
In the $C=+$ scalar channel, coherent Coulomb and instanton attraction produce a
compact $0^{++}$ ground state well below the confinement baseline.
The  tensor $2^{++}$  channel describes a Reggeized orbital sequence. 
The discrepancy observed in the lowest state is due to its sole treatment as
a D-wave, while the reality is more like an S-state with D-wave admixture
as we discuss in the semi-classical analysis. 
By contrast, the vector channels appear only at significantly higher masses and do
not admit compact two-gluon realizations, in agreement with symmetry constraints
and lattice spectroscopy~\cite{Morningstar:1999rf,Morningstar:2025glueballreview}.

%The lattice spectrum shown in Fig.~\ref{fig:glueball_spectrum_PC} is taken from the
%quenched SU(3) calculation of Morningstar and Peardon, including their quoted
%systematic uncertainties.

\begin{table}[b]
\centering
\begin{tabular}{|c|c|c|}
\hline
\bf $J^{PC}$ & \bf This work & \bf Lattice \cite{Meyer:2004gx,Morningstar:1999rf} \\
\hline
$0^{++}$ & 1.92 \quad (BS: 1.4--1.5) & 1.475(30)(65) \\
$0^{++}$ & 2.79  & 2.755(70)(120) \\
$0^{++}$ & 3.29  & 3.370(100)(150) \\
$0^{++}$ & 4.00  & 3.990(210)(180) \\
\hline
$2^{++}$ & 3.01  & 2.150(30)(100) \\
$2^{++}$ & 3.46  & 2.880(100)(130) \\
$2^{++}$ & 4.16  & 3.385(90)(150) \\
$2^{++}$ & 5.10  & -- \\
\hline
$4^{++}$ & 3.52  & 3.640(90)(160) \\
$4^{++}$ & 4.18  & -- \\
$4^{++}$ & 5.10  & -- \\
$4^{++}$ & 6.24  & -- \\
\hline
$6^{++}$ & 4.06  & 4.360(260)(200) \\
$6^{++}$ & 4.97  & -- \\
$6^{++}$ & 6.11  & -- \\
$6^{++}$ & 7.49  & -- \\
\hline
$0^{-+}$ & 2.65  & 2.250(60)(100) \\
$0^{-+}$ & 3.13  & 3.370(150)(150) \\
\hline
$1^{+-}$ & --    & 2.94(17) \\
$1^{--}$ & --    & 3.81(27) \\
\hline
$3^{++}$ & --    & 3.38(15) \\
$3^{+-}$ & --    & 3.70(23) \\
$3^{--}$ & --    & 4.03(25) \\
\hline
\end{tabular}
\caption{
Glueball masses in pure SU(3) gauge theory: comparison of this work as in Fig.~\ref{fig_gbpp}, with lattice results from Refs.~\cite{Morningstar:1999rf,Meyer:2004gx}.
}
\label{tab:glueball_twocol_JPC}
\end{table}

\section{Conclusions}
\label{sec_summary}

We have developed a constituent two-gluon Hamiltonian framework for glueballs that incorporates  adjoint Coulomb interaction, instanton-induced short-range forces, and full tensor-driven S-D mixing in the $2^{++}$ channel, with parameters calibrated directly against quenched lattice Yang-Mills results. Within this framework, several robust conclusions emerge that align naturally with current lattice spectroscopy.

At a qualitative level, the emerging picture is that of a rescaled theory of mesons. The effective gluon mass, $m_g\approx 0.9\, \text{GeV}$, lies between the strange and charm quark masses. The interaction---both perturbative and confining, with the latter derived from adjoint instanton molecules---is enhanced by a factor of $9/4$. The bulk of the spectroscopy then follows from standard solutions of the Schr\"odinger equation, in close analogy with the case of charmonium.

Then there are channels for which
forces are too strong to be treated in this way. The main example is
 the scalar $0^{++}$ glueball is strongly compressed by the combined effect of the attractive adjoint Coulomb interaction and an instanton-induced core attraction. The resulting compact radius, the smallest of all hadrons, mirrors lattice indications.  This behavior arises dynamically and does not require fine tuning, supporting the interpretation of the scalar glueball as a state dominated by short-distance nonperturbative physics.

 The tensor $2^{++}$ glueball remains spatially extended. Centrifugal suppression and substantial S-D wave mixing driven by the tensor interaction reduce short-distance overlap and weaken instanton effects. As a result, the tensor mass is shifted only moderately relative to the confinement baseline, and its radius remains considerably larger than that of the scalar, consistent with lattice mass hierarchies and emerging lattice probes of glueball spatial structure.

The semiclassical WKB analysis provides analytic insight into the lattice-observed organization of glueball excitations. Adjoint confinement fixes the asymptotic Regge slopes, while Coulomb and instanton-induced interactions generate negative, spin-dependent mass shifts that predominantly affect low-$J$ states. This naturally explains the downward curvature of Regge trajectories near the intercept and the rapid decoupling of instanton physics for higher orbital and radial excitations seen in lattice spectra.

Finally, symmetry constraints combined with the structure of instanton-induced interactions imply that vector glueball channels are weakly bound or absent in the two-gluon sector, in agreement with lattice results that place the lightest vector and pseudovector glueballs at substantially higher masses and suggest a dominant multigluon structure.

Overall, the present analysis provides a coherent constituent-gluon interpretation of lattice glueball spectroscopy, linking mass hierarchies, spatial sizes, and Regge behavior to a small set of physically motivated nonperturbative mechanisms. This framework offers a useful bridge between lattice calculations and phenomenological modeling, and can be systematically extended to explore glueball form factors, and mixing with quarkonia in full QCD.\\
%, and the behavior of gluonic bound states at finite temperature or density.

{\centerline{\bf Acknowledgements}}
\vskip 0.5cm
This work is supported by the Office of Science, U.S. Department of Energy under Contract  No. DE-FG-88ER40388.
This research is also supported in part within the framework of the Quark-Gluon Tomography (QGT) Topical Collaboration, under contract no. DE-SC0023646.

\appendix

\section{Gluon mass and density scaling}
\label{app:ME}

Musakhanov and Egamberdiev~\cite{Musakhanov:2017erp} derived the gluon polarization operator in the instanton medium and extracted a momentum-dependent dynamical mass. In their ILM setup the scalar ``gluon'' mass $M_s(q)$ is generated by rescattering on instantons, and the physical transverse gluon mass satisfies $M_g^2(q)=2M_s^2(q)$, implying at $q=0$ the standard-ILM value $M_g(0)\simeq 0.36~\mathrm{GeV}$ \cite{Musakhanov:2017erp}. The key scaling with density follows from the structure of the self-energy in a random instanton background: to leading order in the density expansion the polarization operator is proportional to the instanton density $n$, and the mass is obtained from the infrared limit of the propagator denominator, schematically $D^{-1}(q)\sim q^2+\Pi(q)$ with $\Pi(0)\propto n$. This implies $m_g^2\propto n$ and therefore
\begin{equation}
m_g(\eta)=m_{g0}\sqrt{\eta},
\qquad \eta\equiv \frac{n_{\rm eff}}{n_0},
\label{eq:mgScalingApp}
\end{equation}
where $n_0\simeq 1~\mathrm{fm}^{-4}$ is the traditional ILM density used to define $m_{g0}$. In the dense-ILM-ensemble~\cite{Shuryak:1982ilm}, $\eta$ is naturally taken as a 
limit of gradient 
flow time going to zero.
It is relatively uncertain 
from original studies, we use
$\eta=7$.

%++++++++++++++++NEW V

%\begin{table}[h]
%\centering
%\begin{tabular}{@{}cccccc@{}}
%\toprule
%$n$ & $L$ & $J$ (trajectory) & $M_{nL}^{\rm WKB}$ [GeV] & $r_+$ [fm] & $r_{\rm rms}^{\rm WKB}$ [fm] \\
%\midrule
%0 & 0 & 0 & 2.505 & 0.870 & 0.635 \\
%1 & 0 & 0 & 3.861 & 1.530 & 1.118 \\
%2 & 0 & 0 & 4.965 & 2.068 & 1.511 \\
%0 & 2 & 2 & 3.861 & 1.530 & 1.118 \\
%1 & 2 & 2 & 4.965 & 2.068 & 1.511 \\
%0 & 4 & 4 & 4.965 & 2.068 & 1.511 \\
%0 & 6 & 6 & 5.941 & 2.544 & 1.858 \\
%\bottomrule
%\end{tabular}
%\caption{WKB masses and radii at $\eta=1$ for the linear adjoint potential. Here $J$ is identified with $L$ for Regge-trajectory purposes. These semiclassical radii are larger than the variational scalar radius because Coulomb and instanton attractions are not included in the baseline WKB table; they can be incorporated as short-distance energy shifts (see Appendix~\ref{app:ILMpot}).}
%\label{tab:wkb}
%\end{table}

%++++++++++++++++++++++++NEW VI

\begin{table*}[t]
\centering
\begin{tabular}{cccccc}
\hline\hline
$J$ &
$M^{\rm WKB}$ (GeV) &
$\Delta M(J)$ (GeV) &
$M^{\rm WKB}+\Delta M$ (GeV) &
$(M^{\rm WKB})^2$ (GeV$^2$) &
$(M^{\rm WKB}+\Delta M)^2$ (GeV$^2$) \\
\hline
0 & 3.196 & $-1.466$ & 1.730 & 10.213 & 2.993 \\
2 & 4.176 & $-0.681$ & 3.495 & 17.438 & 12.215 \\
4 & 4.974 & $-0.481$ & 4.493 & 24.745 & 20.190 \\
6 & 5.680 & $-0.383$ & 5.296 & 32.257 & 28.049 \\
\hline\hline
\end{tabular}
\caption{Smoothly shifted trajectory points at $\eta=7$. The unshifted WKB baseline is computed from adjoint linear confinement. The shift is anchored at the physical scalar intercept $M_{0^{++}}(J{=}0)=1.73~{\rm GeV}$ and decreases smoothly with $J$.}
\label{tab:wkb_tableIII}
\end{table*}

\section{Semiclassical (WKB) spectrum}
\label{sec:semi}
A complementary analytic description follows from the semiclassical quantization. For the linear potential $V(r)=\sigma_8 r$ with reduced mass $\mu=m_g/2$, the WKB spectrum with the Langer modification yields
\begin{eqnarray}
E_{nL}^{\rm WKB}(\eta)&=&
\left[\frac{3\pi}{2}\left(n+\frac{L}{2}+\frac{3}{4}\right)\right]^{2/3}
\frac{\sigma_8^{2/3}}{m_g^{1/3}(\eta)},
\nonumber\\
M_{nL}^{\rm WKB}(\eta)&=&2m_g(\eta)+E_{nL}^{\rm WKB}(\eta),
\label{eq:WKBspec}
\end{eqnarray}
which implies
\begin{equation}
M_{nJ}^2
\simeq
\frac{9\pi^2}{4}\,
\frac{\sigma_8^{4/3}}{m_g^{2/3}}\,
J^{4/3}
\qquad
(J\gg1).
\end{equation}
The  nonrelativistic WKB Hamiltonian with a linear potential produces an asymptotic power-law ``quasi-Regge'' behavior $M^2\propto J^{4/3}$ rather than a strictly linear $M^2\propto J$. Over the moderate range of spins accessible in typical lattice spectra, the resulting $M^2(J)$ are approximately linear.

%%%%%%%%

In Table~\ref{tab:wkb_tableIII} we list the unshifted WKB baseline  for the $0^{++}$ trajectory, computed from the adjoint linear confinement with $\eta=7$. The shift is anchored at the physical scalar intercept $M_{0^{++}}(J{=}0)=1.73~{\rm GeV}$, and decreases smoothly with $J$. In Table~\ref{tab:wkb_tableIV} we also list the unshifted $2^{++}$ WKB baseline with $\eta=7$. 
 The smooth shift $\Delta M_T(J)$ is anchored at the physical tensor point $M_{2^{++}}(J{=}2)=2.40~{\rm GeV}$,  and decreases with $J$ as the short-distance overlap is suppressed by increasing orbital angular momentum and by $\,{}^{5}\!S_2$-$\,{}^{5}\!D_2$ mixing. In both tables, the deformation is numerically anchored at the lowest state, and smoothly vanishes asymptotically.

The Regge trajectories discussed so far separate naturally into two components:
an asymptotic contribution governed by adjoint confinement and a low-$J$ deformation
generated by short-distance dynamics. While the former fixes the Regge slope,
the latter controls the intercept and the curvature of the trajectory near $J=0$.
%We analyze the origin, sign, and $J$-dependence of the deformation
%in the present constituent two-gluon framework.

%++++++++++++++++++NEW VII

\begin{table*}[t]
\centering
%\caption{Tensor-family smooth-shift spectrum at $\eta=7$.}
\begin{tabular}{cccccc}
\hline\hline
$J$ &
$M^{\rm WKB}$ (GeV) &
$\Delta M_T(J)$ (GeV) &
$M^{\rm WKB}+\Delta M_T$ (GeV) &
$(M^{\rm WKB})^2$ (GeV$^2$) &
$(M^{\rm WKB}+\Delta M_T)^2$ (GeV$^2$) \\
\hline
2 & 4.176 & $-1.776$ & 2.400 & 17.438 & 5.760 \\
4 & 4.974 & $-1.214$ & 3.760 & 24.745 & 14.139 \\
6 & 5.680 & $-0.955$ & 4.725 & 32.257 & 22.324 \\
8 & 6.324 & $-0.800$ & 5.524 & 39.990 & 30.509 \\
\hline\hline
\end{tabular}
\caption{Tensor-family smooth-shift trajectory points at $\eta=7$.
%used in Fig.~2. 
The unshifted WKB baseline is computed from adjoint linear confinement. The smooth shift $\Delta M_T(J)$ is anchored at the physical tensor point $M_{2^{++}}(J{=}2)=2.40~{\rm GeV}$, and decreases with $J$ as short-distance overlap is suppressed by increasing orbital angular momentum and by $\,{}^{5}\!S_2$-$\,{}^{5}\!D_2$ mixing.}
\label{tab:wkb_tableIV}
\end{table*}

\begin{figure}[t!]
    \centering   \includegraphics[width=0.85\linewidth]{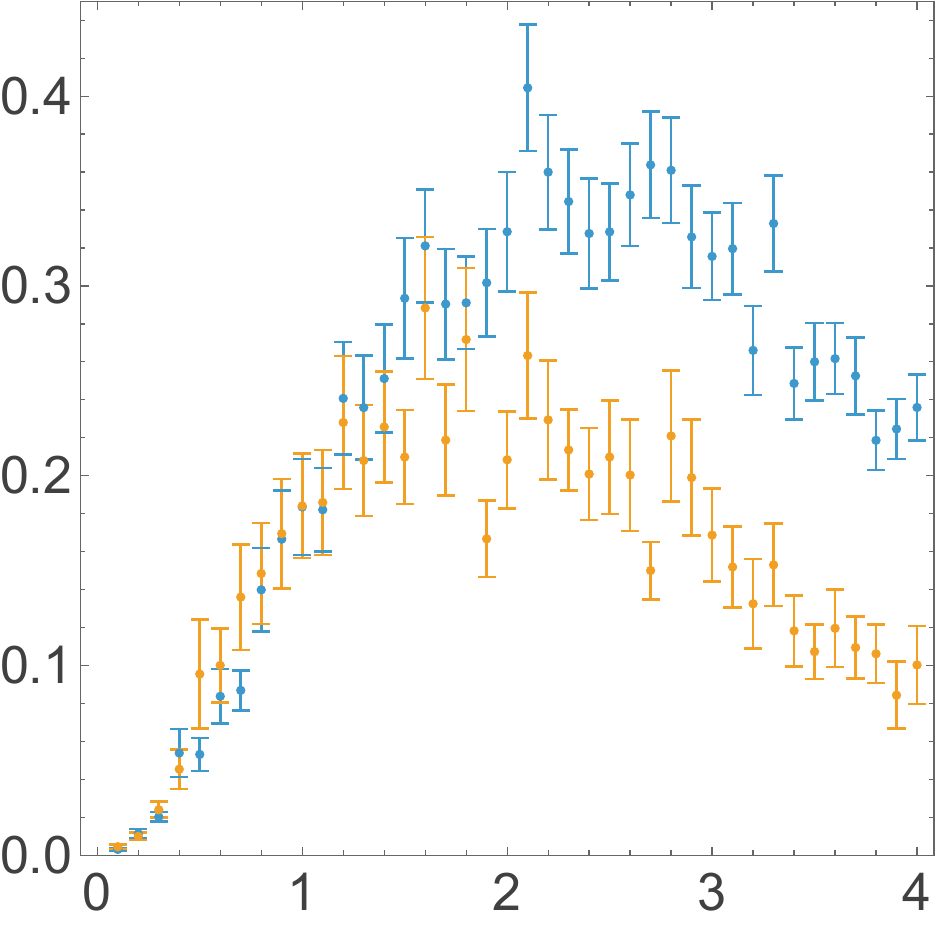}
    \includegraphics[width=0.85\linewidth]{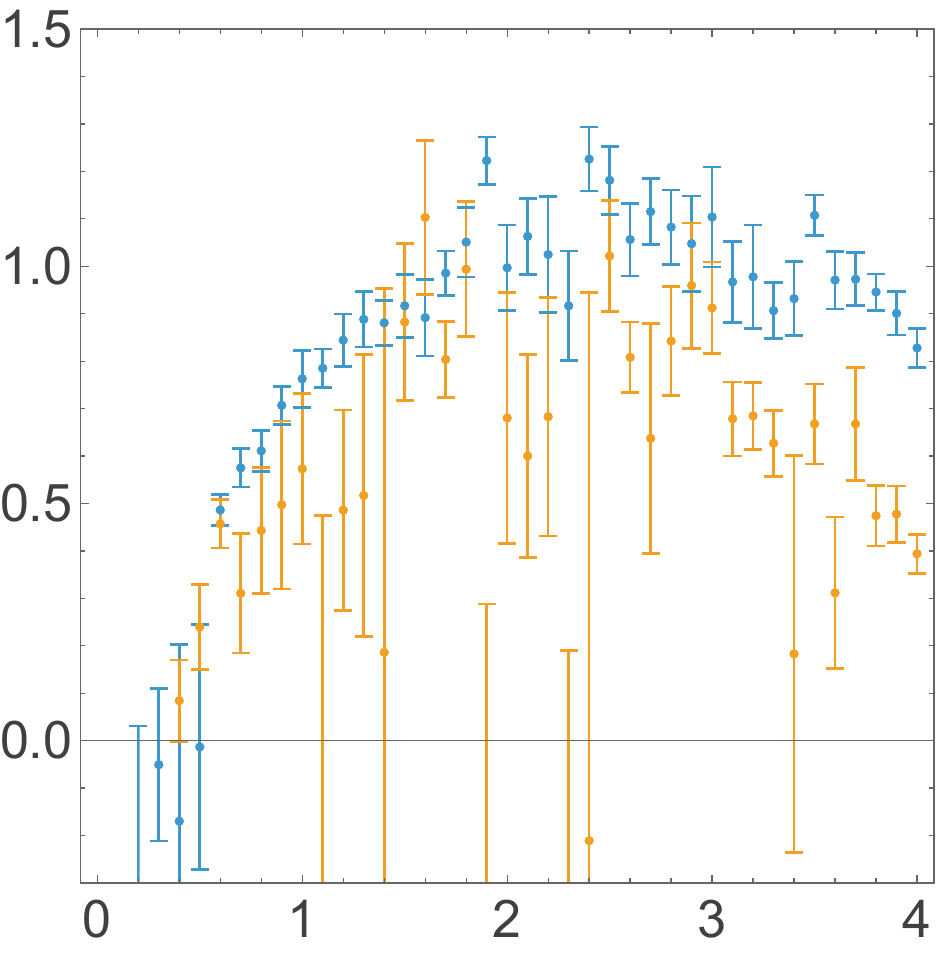}
    \caption{Static potentials $V_{\rm conf}(r)\, (\rm GeV)$ versus $r\, (\rm GeV^{-1})$ calculated for quarks (fundamental representation, upper plot) and gluons (adjoint representation, lower plot )
    Wilson lines. Blue dots are for a  $\bar I I$ molecule, red are for a single instanton. Potentials still needs to be rescaled proportional to their density.}
\label{fig_W_fund_adj}
\end{figure}

A convenient WKB size measure is the outer turning point $r_+=E/\sigma_8$. For $L=0$, the WKB radial probability density implies closed expressions for moments. With $p(r)=\sqrt{2\mu(E-\sigma_8 r)}$ and 
$$|\psi|^2 d^3r\propto P(r)dr \propto dr/p(r)\,,$$ 
hence
\begin{equation}
P(r)=\frac{1}{2\sqrt{r_+}}\frac{1}{\sqrt{r_+-r}},
\qquad r_+=\frac{E}{\sigma_8},
\label{eq:Pr}
\end{equation}
In particular, we have
\begin{eqnarray}
\langle r\rangle_{L=0}^{\rm WKB}&=&\frac{2}{3}r_+,
\nonumber\\
\langle r^2\rangle_{L=0}^{\rm WKB}&=&\frac{8}{15}r_+^2,
\nonumber\\
r_{{\rm rms},\,L=0}^{\rm WKB}&=&\sqrt{\frac{8}{15}}\,r_+.
\label{eq:WKBrms}
\end{eqnarray}
 The moments follow from Beta-function integrals, giving \eqref{eq:WKBrms}. The same turning point definition applies at general $(n,L)$ with $r_+=E_{nL}^{\rm WKB}/\sigma_8$, and provides an effective size scale for the excited states and Regge trajectories.
For $L>0$ we quote $r_+$ and use $r_{\rm rms}\approx \sqrt{8/15}\,r_+$ as a uniform semiclassical estimate, which is adequate for trajectory comparisons.

Table~\ref{tab:wkb} provides the WKB masses and radii at $\eta=7$ for low-lying $L=0$ (scalar-like) and $L=2$ (tensor-like orbital) levels, including the first few excitations. The conversion $1~\mathrm{GeV}^{-1}=0.197~\mathrm{fm}$ is used.

\begin{table}[ht]
\centering
%\caption{WKB masses and radii for two–gluon glueballs at $\eta=7$.}
\begin{tabular}{cccccc}
\hline\hline
$n$ & $L$ & $J$ & $M^{\rm WKB}_{nL}$ (GeV) & $r_+$ (fm) & $r^{\rm WKB}_{\rm rms}$ (fm) \\
\hline
0 & 0 & 0 & 3.196 & 0.628 & 0.459 \\
1 & 0 & 0 & 4.176 & 1.105 & 0.807 \\
2 & 0 & 0 & 4.974 & 1.493 & 1.090 \\
0 & 2 & 2 & 4.176 & 1.105 & 0.807 \\
1 & 2 & 2 & 4.974 & 1.493 & 1.090 \\
0 & 4 & 4 & 4.974 & 1.493 & 1.090 \\
0 & 6 & 6 & 5.680 & 1.836 & 1.341 \\
\hline\hline
\end{tabular}
\caption{WKB masses and radii at $\eta=7$ for the linear adjoint potential. Here $J$ is identified with $L$ for Regge-trajectory purposes. These semiclassical radii are larger than the variational scalar radius because Coulomb and instanton attractions are not included in the baseline WKB table; they can be incorporated as short-distance energy shifts.} %(see Appendix~\ref{app:ILMpot}).}
\label{tab:wkb}
\end{table}

\section{Static potentials for fundamental and adjoint charges from instantons and $I\bar I$ molecules}
\label{sec_pote_numeric}

Wilson loops involve path-ordered exponents
\be
W_C = P \exp\!\left[i \sum \Delta x_\mu\, A_\mu^a\, T^a \right],
\ee
taken over closed contours $C$, usually of rectangular
shape. The sum runs over infinitesimal elements
$\Delta x_\mu$ along the loop, and $A_\mu^a$ are vacuum
gauge fields. Since the contour is closed, $W_C$ is
gauge invariant.

For static color charges corresponding to quarks,
the color generators are in the fundamental
$SU(N_c)$ representation, $T^a=\lambda^a/2$, with
Pauli or Gell-Mann matrices. In several of our
earlier publications we have numerically calculated
the corresponding static confining potentials from
ensembles of instantons or $I\bar I$ molecules, and
applied them to quarkonia, baryons, and tetraquarks
\cite{}. The upper panel of Fig.~\ref{fig_W_fund_adj}
shows an example of such a Monte-Carlo simulation.
(The Wilson line locations and orientations are
randomized.)

Since a ``constituent gluon'' belongs to the adjoint
color representation, the only modification in the
Wilson loop is that the generators are now adjoint,
\be
(T^a)_{bc} = -i\,\epsilon^{a}_{\ bc}.
\ee
Unlike the case of Pauli matrices, for which a
compact closed form for the exponent exists, no such
expression is available here. The practical method
we employ is based on a Taylor expansion of the
exponent to second order in $\Delta x_\mu$, assumed
to be small. The adjoint potential shown in the lower
panel of Fig.~\ref{fig_W_fund_adj} was calculated in
this way, using the same gauge-field configurations
and the same set of Wilson lines as in the upper
panel.

First, note that the vertical scales of the two plots
are different, and that, crudely, the ratio of the
two potentials is approximately $9/4$, as suggested
by the ratio of the corresponding color Casimir
operators.

Second, since the Wilson lines and the resulting
potentials contain nonlinear terms (higher powers of
the gauge field), such scaling is not expected to
hold exactly. Indeed, the shapes of the two
potentials differ somewhat. Nevertheless, all
potentials reach a maximum at
$r \sim 2.5\,\mathrm{GeV}^{-1} \sim 0.5\,\mathrm{fm}$
and then decrease slightly. This feature is likely
an artifact of the setup used here, in which a single
instanton (or molecule) is surrounded by a Wilson
loop. In an ensemble of such objects, the potential
is expected to saturate at large $r$ to a constant,
equal to twice the effective mass of the static
charges.

For heavy quarks interacting via instantons, this
asymptotic value is
\be
V_{\rm conf}(r \rightarrow \infty)
\approx 2\,\eta \times (0.070\,\mathrm{GeV})
\approx 1\,\mathrm{GeV},
\ee
for $\eta \sim 7$, corresponding to a rescaling from
the dilute instanton liquid model to a dense ensemble
of molecules. In physical QCD such an energy is
sufficient to produce an additional $\bar q q$ pair
and split quarkonium $\bar Q Q$ into two $\bar Q q$
mesons.

For ``constituent gluons'' in pure gauge theory, the
large-distance splitting instead corresponds to
$gg \rightarrow (gg)(gg)$, so that
$V_{\rm conf}(r \rightarrow \infty)$ must be at least
of order $M_{0^{++}} \sim 1.5\,\mathrm{GeV}$. Guided by
these considerations, we model our potential in the
form shown in Fig.~\ref{fig_pot}.

\section{Details of the ${}^5S_2$-${}^5D_2$ mixing in the tensor channel}
\label{app:SDmix}

For two spin-1 constituents, the tensor glueball $2^{++}$ is dominated by total spin $S=2$ and total $J=2$, with orbital components $L=0$ and $L=2$ mixed by the tensor operator $S_{12}$. In the coupled basis $\{\ket{{}^5S_2},\ket{{}^5D_2}\}$, the tensor operator has the standard reduced matrix elements
\begin{widetext}
\begin{equation}
\langle {}^5S_2|S_{12}|{}^5S_2\rangle=0,
\qquad
\langle {}^5D_2|S_{12}|{}^5D_2\rangle=-\frac{2}{7},
\qquad
\langle {}^5S_2|S_{12}|{}^5D_2\rangle=\sqrt{\frac{8}{7}},
\label{eq:S12matrix}
\end{equation}
up to phase conventions.

Using the Gaussian trial functions \eqref{eq:psi_S} and \eqref{eq:psi_D}, the $D$-wave normalization is
\begin{equation}
\mathcal N_D=\left(\frac{8}{15}\right)^{1/2}\left(\frac{\beta_D^2}{\pi}\right)^{3/4}\beta_D^2,
\label{eq:ND}
\end{equation}
and the $D$-wave kinetic and linear moments are
\begin{equation}
\langle \bm p^2\rangle_D=\frac{7}{2}\beta_D^2,
\qquad
\langle r\rangle_D=\frac{16}{3\sqrt{\pi}}\frac{1}{\beta_D},
\qquad
\Big\langle e^{-r^2/\rho^2}\Big\rangle_D=\left(\frac{\beta_D^2}{\beta_D^2+\rho^{-2}}\right)^{7/2}.
\label{eq:Dexpvals}
\end{equation}
The corresponding spin-independent functional is
\begin{align}
E_D^{(0)}(\beta_D;\eta)=2m_g(\eta)+\frac{7\beta_D^2}{2m_g(\eta)}
+\sigma_8\frac{16}{3\sqrt{\pi}\beta_D}
-\frac{6\alpha_s^{\rm eff}}{\sqrt{\pi}}\beta_D\,\kappa_D
-G(\eta)\left(\frac{\beta_D^2}{\beta_D^2+\rho^{-2}}\right)^{7/2},
\label{eq:ED0}
\end{align}
where $\kappa_D$ encodes the reduced $D$-wave expectation of $1/r$ relative to the $S$-wave (it is an $\mathcal O(1)$ number obtained by a straightforward radial integral; for compactness we keep it symbolic here since the tensor state is typically controlled more by confinement and mixing than by the Coulomb core).

The tensor-sector $2\times2$ Hamiltonian is
\begin{equation}
\mathbb H_{J=2}=
\begin{pmatrix}
E_S^{(0)}(\beta_S;\eta)+\Delta_{SS}^{(S=2)}(\eta) & V_{SD}(\eta)\\[4pt]
V_{SD}(\eta) & E_D^{(0)}(\beta_D;\eta)+\Delta_{SS,D}^{(S=2)}(\eta)+\Delta_{T,DD}(\eta)
\end{pmatrix},
\label{eq:H2x2}
\end{equation}
with
\begin{equation}
\Delta_{SS}^{(S=2)}(\eta)=\frac{C_{SS}(\eta)}{m_g^2(\eta)}\langle \delta_\Lambda^{(3)}\rangle_S,
\qquad
\Delta_{SS,D}^{(S=2)}(\eta)=\frac{C_{SS}(\eta)}{m_g^2(\eta)}\langle \delta_\Lambda^{(3)}\rangle_D,
\label{eq:SSshifts}
\end{equation}
and tensor terms
\begin{equation}
V_{SD}(\eta)=\frac{C_T(\eta)}{m_g^2(\eta)}\,\sqrt{\frac{8}{7}}\,
\int_0^\infty dr\,r^2\,R_S(r)\,R_D(r)\,\frac{1-e^{-r^2/\rho^2}}{r^3},
\label{eq:VSDint}
\end{equation}
\begin{equation}
\Delta_{T,DD}(\eta)=\frac{C_T(\eta)}{m_g^2(\eta)}\left(-\frac{2}{7}\right)\,
\int_0^\infty dr\,r^2\,|R_D(r)|^2\,\frac{1-e^{-r^2/\rho^2}}{r^3}.
\label{eq:VDDint}
\end{equation}
\end{widetext}
Diagonalizing \eqref{eq:H2x2} gives the physical tensor mass as the lower eigenvalue and defines a mixing angle $\theta$ by
\begin{equation}
\tan 2\theta=\frac{2V_{SD}}{H_{DD}-H_{SS}}.
\label{eq:theta}
\end{equation}
The tensor rms radius is then computed from the mixed state,
\begin{equation}
\langle r^2\rangle_{2^{++}}=\cos^2\theta\,\langle r^2\rangle_S+\sin^2\theta\,\langle r^2\rangle_D,
%\qquad
%r_{2^{++}}=\sqrt{\langle r^2\rangle_{2^{++}}}.
\label{eq:r2mix}
\end{equation}
with $r_{2^{++}}=\sqrt{\langle r^2\rangle_{2^{++}}}$.
The interference term vanishes because $r^2$ is purely radial and orthogonal between $L=0$ and $L=2$.

\section{Details of the radial integrations for tensor mixing}
\label{app:SDmix_full}

This appendix provides the explicit analytic evaluation of all radial integrals entering the
${}^5S_2$-${}^5D_2$ tensor mixing problem and the corresponding expectation values of
the Hamiltonian terms.

\subsection{Normalized trial wavefunctions}
For simplicity, we will use  variational estimates of matrix elements using simplified wave functions.
The $S$-wave Gaussian trial function is
\begin{equation}
\psi_S(\bm r)
=
\left(\frac{\beta_S^2}{\pi}\right)^{3/4}
e^{-\beta_S^2 r^2/2},
\label{eq:psi_S}
\end{equation}
with radial part
\begin{equation}
R_S(r)
=
\left(\frac{2\beta_S^3}{\sqrt{\pi}}\right)^{1/2}
e^{-\beta_S^2 r^2/2}.
\end{equation}

The $D$-wave trial function is
\begin{equation}
\psi_{D,m}(\bm r)
=
\mathcal N_D\,
r^2 e^{-\beta_D^2 r^2/2}
Y_{2m}(\hat{\bm r}),
\label{eq:psi_D}
\end{equation}
with normalization fixed by
\begin{equation}
1
=
\int d^3r\,|\psi_{D,m}(\bm r)|^2
=
\mathcal N_D^2
\int_0^\infty dr\, r^6 e^{-\beta_D^2 r^2},
\end{equation}
which yields
\begin{equation}
\mathcal N_D
=
\left(\frac{8}{15}\right)^{1/2}
\left(\frac{\beta_D^2}{\pi}\right)^{3/4}
\beta_D^2.
\end{equation}
The radial function is therefore
\begin{equation}
R_D(r)
=
\left(\frac{8}{15}\right)^{1/2}
\left(\frac{\beta_D^2}{\pi}\right)^{3/4}
\beta_D^2\,
r^2 e^{-\beta_D^2 r^2/2}.
\end{equation}

\subsection{Kinetic and confining expectation values}

Using standard Gaussian integrals,
\begin{equation}
\langle \bm p^2\rangle_S
=
\frac{3}{2}\beta_S^2,
\qquad
\langle \bm p^2\rangle_D
=
\frac{7}{2}\beta_D^2.
\end{equation}

The linear confinement expectation values are
\begin{equation}
\langle r\rangle_S
=
\frac{2}{\sqrt{\pi}}\frac{1}{\beta_S},
\qquad
\langle r\rangle_D
=
\frac{16}{3\sqrt{\pi}}\frac{1}{\beta_D}.
\end{equation}

The corresponding rms radii are
\begin{equation}
\langle r^2\rangle_S
=
\frac{3}{2\beta_S^2},
\qquad
\langle r^2\rangle_D
=
\frac{7}{2\beta_D^2}.
\end{equation}

\subsection{Coulomb expectation values}

For the Coulomb potential $V_C(r)=-3\alpha_s^{\rm eff}/r$, one finds
\begin{equation}
\left\langle \frac{1}{r}\right\rangle_S
=
\frac{2\beta_S}{\sqrt{\pi}},
\end{equation}
and for the $D$-wave,
\begin{equation}
\left\langle \frac{1}{r}\right\rangle_D
=
\frac{8}{3\sqrt{\pi}}\beta_D.
\end{equation}
Thus the Coulomb term is parametrically suppressed in the tensor channel by angular momentum.

\subsection{Instanton central attraction}

The scalar instanton term yields
\begin{equation}
\left\langle e^{-r^2/\rho^2}\right\rangle_S
=
\left(\frac{\beta_S^2}{\beta_S^2+\rho^{-2}}\right)^{3/2}.
\end{equation}
For the $D$-wave,
\begin{equation}
\left\langle e^{-r^2/\rho^2}\right\rangle_D
=
\left(\frac{\beta_D^2}{\beta_D^2+\rho^{-2}}\right)^{7/2},
\end{equation}
showing that instanton attraction is strongly suppressed for higher partial waves.

\subsection{Spin-spin contact term}

With the Gaussian regulator
\begin{equation}
\delta^{(3)}_\Lambda(\bm r)
=
\left(\frac{\Lambda^2}{\pi}\right)^{3/2}
e^{-\Lambda^2 r^2},
\end{equation}
the contact expectation values are
\begin{equation}
\langle \delta_\Lambda^{(3)}\rangle_S
=
\left(\frac{\Lambda^2}{\Lambda^2+\beta_S^2}\right)^{3/2}
\left(\frac{\beta_S^2}{\pi}\right)^{3/2},
\label{eq:deltagammaS}
\end{equation}
and
\begin{equation}
\langle \delta_\Lambda^{(3)}\rangle_D
=
\left(\frac{\Lambda^2}{\Lambda^2+\beta_D^2}\right)^{7/2}
\left(\frac{\beta_D^2}{\pi}\right)^{3/2}
\frac{15}{8}.
\end{equation}

\subsection{Tensor matrix elements}

The tensor operator enters through
\begin{equation}
V_T(r)=\frac{C_T(\eta)}{m_g^2(\eta)}\,
\frac{1-e^{-r^2/\rho^2}}{r^3} S_{12}.
\end{equation}
The reduced angular matrix elements in the
$\{{}^5S_2,{}^5D_2\}$ basis are
\begin{widetext}
\begin{equation}
\langle {}^5S_2|S_{12}|{}^5S_2\rangle=0,
\quad
\langle {}^5D_2|S_{12}|{}^5D_2\rangle=-\frac{2}{7},
\quad
\langle {}^5S_2|S_{12}|{}^5D_2\rangle=\sqrt{\frac{8}{7}}.
\end{equation}
\end{widetext}

The radial mixing integral is
\begin{equation}
I_{SD}
=
\int_0^\infty dr\,
R_S(r)R_D(r)\,
\frac{1-e^{-r^2/\rho^2}}{r}.
\end{equation}
which yields
\begin{equation}
I_{SD}
=
\mathcal C
\left[
\frac{1}{(\beta_S^2+\beta_D^2)^{1/2}}
-
\frac{1}{(\beta_S^2+\beta_D^2+\rho^{-2})^{1/2}}
\right],
\label{eq:isd}
\end{equation}
with the normalization factor
\begin{equation}
\mathcal C
=
\left(\frac{16}{15\pi}\right)^{1/2}
\beta_S^{3/2}\beta_D^{7/2}.
\end{equation}

The diagonal $D$-wave tensor term is
\begin{equation}
I_{DD}
=
\int_0^\infty dr\,
R_D^2(r)\,
\frac{1-e^{-r^2/\rho^2}}{r},
\end{equation}
which yields
\begin{equation}
I_{DD}
=
\frac{4}{7}
\left[
\beta_D
-
\frac{\beta_D^2}{(\beta_D^2+\rho^{-2})^{1/2}}
\right].
\label{eq:idd}
\end{equation}

%The $2\times2$ tensor Hamiltonian matrix elements follow directly from these expressions.

\subsection{Explicit ${}^5S_2$-${}^5D_2$ tensor Hamiltonian}
\label{app:SDmix_matrix}

In the coupled basis $\{|{}^5S_2\rangle, |{}^5D_2\rangle\}$ the tensor glueball Hamiltonian
takes the explicit $2\times2$ form
\begin{equation}
\mathbb{H}_{2^{++}}(\eta)=
\begin{pmatrix}
H_{SS}(\eta) & H_{SD}(\eta)\\[6pt]
H_{SD}(\eta) & H_{DD}(\eta)
\end{pmatrix},
\label{eq:H2explicit}
\end{equation}
with the diagonal $S$-wave entry
\begin{equation}
H_{SS}(\eta)
=
E_S^{(0)}(\beta_S;\eta)
+
\frac{C_{SS}(\eta)}{m_g^2(\eta)}
\left\langle \delta^{(3)}_\Lambda \right\rangle_S ,
\label{eq:HSS}
\end{equation}
where $\langle \delta^{(3)}_\Lambda\rangle_S$ is given in
Eq.~\eqref{eq:deltagammaS}.

The diagonal $D$-wave entry reads
\begin{eqnarray}
&&H_{DD}(\eta)
=
E_D^{(0)}(\beta_D;\eta)\nonumber\\
&&+
\frac{C_{SS}(\eta)}{m_g^2(\eta)}
\left\langle \delta^{(3)}_\Lambda \right\rangle_D
-
\frac{2}{7}\,
\frac{C_T(\eta)}{m_g^2(\eta)}\,
I_{DD},\nonumber\\
\label{eq:HDD}
\end{eqnarray}
where the factor $-2/7$ is the reduced angular matrix element
$\langle{}^5D_2|S_{12}|{}^5D_2\rangle$ and $I_{DD}$ is given  in Eq.~\eqref{eq:idd}.

The off-diagonal mixing term is
\begin{eqnarray}
H_{SD}(\eta)
=
\sqrt{\frac{8}{7}}\,
\frac{C_T(\eta)}{m_g^2(\eta)}\,
I_{SD},
\label{eq:HSD}   
\end{eqnarray}
where $\sqrt{8/7}$ is the reduced matrix element
$\langle{}^5S_2|S_{12}|{}^5D_2\rangle$ and $I_{SD}$ is given in
Eq.~\eqref{eq:isd}.

The physical tensor glueball mass is the lower eigenvalue
\begin{eqnarray}
&&M_{2^{++}}(\eta)
=
\frac{1}{2}\left(H_{SS}+H_{DD}\right)\nonumber\\
&&-
\frac{1}{2}
\sqrt{\left(H_{SS}-H_{DD}\right)^2+4H_{SD}^2},
\label{eq:M2phys}   
\end{eqnarray}
and the $S$-$D$ mixing angle $\theta$ is defined by
\begin{equation}
\tan 2\theta(\eta)=\frac{2H_{SD}(\eta)}{H_{DD}(\eta)-H_{SS}(\eta)}.
\label{eq:thetaExplicit}
\end{equation}

The tensor rms radius follows directly as
\begin{eqnarray}
\langle r^2\rangle_{2^{++}}
&=&
\cos^2\theta\,\langle r^2\rangle_S
+
\sin^2\theta\,\langle r^2\rangle_D,
\nonumber\\
r_{2^{++}}&=&\sqrt{\langle r^2\rangle_{2^{++}}}.
\label{eq:rTensorFinal} 
\end{eqnarray}

\section{Model parameters}
\label{app:params}

This appendix collects all parameters appearing in the Hamiltonian and explains their physical origin and how they enter the calculations.

The fundamental string tension in the fundamental representation is taken as
\begin{equation}
\sigma_3 = 0.18~\mathrm{GeV}^2,
\end{equation}
consistent with lattice determinations in pure Yang-Mills theory. Casimir scaling is assumed for adjoint sources, yielding
\begin{equation}
\sigma_8 = \frac{C_A}{C_F}\sigma_3 = \frac{9}{4}\sigma_3 \simeq 0.405~\mathrm{GeV}^2.
\end{equation}
This parameter enters the linear confining potential and controls the Regge slopes and large-radius behavior.

%The adjoint screening length is taken in the range
%\begin{equation}
%r_{\rm scr}=0.2\text{-}0.4~\mathrm{fm},
%\end{equation}
%motivated by adjoint gluonic  screening. In the variational ground states, the scalar typically probes distances smaller than $r_{\rm scr}$, so the precise value affects mainly excited and tensor states.

The instanton size is fixed at
\begin{equation}
\rho = \frac13~\mathrm{fm} \simeq 1.67~\mathrm{GeV}^{-1},
\end{equation}
as determined phenomenologically and supported by lattice measurements of the instanton size distribution. This parameter sets the range of the instanton-induced interactions and the regulator scale for spin-dependent forces.

The effective gluon mass at unit density is
\begin{equation}
m_{g0} \simeq 0.36~\mathrm{GeV},
\end{equation}
taken from the infrared limit of the gluon propagator in the instanton vacuum. The density scaling
\begin{equation}
m_g(\eta)=m_{g0}\sqrt{\eta}
\end{equation}
reflects the proportionality of the polarization operator to the instanton density~\cite{Musakhanov:2017erp}.

The effective Coulomb coupling is treated phenomenologically,
\begin{equation}
\alpha_s^{\rm eff}=0.35\text{-}0.45,
\end{equation}
corresponding to a moderately strong coupling at distances of order $0.2$-$0.5~\mathrm{fm}$. Its value is constrained by the requirement of a compact scalar glueball without destabilizing the tensor channel.

The instanton-induced scalar attraction strength is parametrized as
\begin{equation}
G(\eta)=G_0\,\eta,
\end{equation}
with $G_0$ fixed at $\eta=1$ to reproduce the lattice scalar glueball mass together with the Coulomb term. Typical fitted values are $G_0\simeq2$-$2.5~\mathrm{GeV}$.

The spin-spin coupling is written as
\begin{equation}
C_{SS}(\eta)=C_{SS}^{(0)}\eta,
\end{equation}
with $C_{SS}^{(0)}$ adjusted to reproduce the $0^{++}$-$2^{++}$ splitting at $\eta=1$. This term primarily controls the relative placement of the scalar and tensor levels and has little effect on radii.

The tensor coupling
\begin{equation}
C_T(\eta)=C_T^{(0)}\eta
\end{equation}
governs $S$-$D$ mixing in the tensor channel. Its sign is allowed to be negative, consistent with dense instanton ensemble estimates of tensor matrix elements. Its magnitude controls the amount of $D$-wave admixture and therefore the tensor glueball radius.

The contact regulator scale is set by the instanton size,
\begin{equation}
\Lambda \simeq \rho^{-1},
\end{equation}
ensuring that spin-dependent forces probe only distances smaller than the instanton core and do not interfere with confinement physics.

All variational calculations minimize the energy with respect to the width parameters $\beta_S$ and $\beta_D$ independently at fixed $\eta$, using the full spin-independent Hamiltonian. Spin-dependent interactions are then evaluated on the optimized wavefunctions, and the tensor sector is diagonalized exactly in the $\{{}^5S_2,{}^5D_2\}$ basis. The WKB analysis uses the same $\sigma_8$ and $m_g(\eta)$ but omits Coulomb and instanton terms at leading order to preserve analytic transparency; their effect is to shift intercepts without altering Regge slopes.

%\subsection{Numerical values of spin-dependent couplings}
%\label{app:spinparams}

%In the numerical analysis presented in the main text and Tables~\ref{tab:var} and \ref{tab:wkb},
%the spin-dependent couplings are fixed at unit density $\eta=1$ by matching the
%pure-gauge lattice masses of the lowest scalar and tensor glueballs
%$M_{0^{++}}\simeq1.73~\mathrm{GeV}$ and $M_{2^{++}}\simeq2.40~\mathrm{GeV}$.

The spin-spin interaction produces the dominant splitting between the scalar and tensor.
Using the optimized $S$-wave width $\beta_S(\eta=1)$ and the regulated contact expectation
value derived in Appendix~\ref{app:SDmix_full}, the required coupling is
\begin{equation}
C_{SS}^{(0)} = 3.0 .
\label{eq:CSSvalue}
\end{equation}
With the density scaling
\begin{equation}
C_{SS}(\eta)=C_{SS}^{(0)}\,\eta ,
\end{equation}
this value reproduces the observed $0^{++}$-$2^{++}$ mass splitting across the
range $0.8\le\eta\le1.2$ used in the tables.

The tensor coupling controls $S$-$D$ mixing and the tensor radius.
The analysis in~\cite{Shuryak:2021yif} indicates
a negative instanton-induced tensor contribution partially cancelling the perturbative
one, we choose
\begin{equation}
C_T^{(0)} = -1.5 ,
\label{eq:CTvalue}
\end{equation}
corresponding to approximately one-half the magnitude of the spin-spin coupling
with opposite sign. The density scaling
\begin{equation}
C_T(\eta)=C_T^{(0)}\,\eta
\end{equation}
ensures that the tensor force grows with the effective instanton density.

With these values, the tensor glueball acquires a moderate $D$-wave admixture
($\sin^2\theta\simeq0.15$ at $\eta=1$), which increases its radius while leaving
its mass within lattice uncertainties.

%++++++++++++++++++++++++++++++++++++++++++++

\section{$0^{++}, 2^{++}$ BS-kernels from ILM}
\label{sec:2++}
The instanton-induced local operator derived in~\ref{sec:relativistic},  may be written schematically as
\begin{equation}
\Delta\mathcal L_{4g}(x)\;\propto\;
\Big(G^a_{\mu\nu}(x)G^{a}_{\mu\nu}(x)\Big)\,
\Big(G^b_{\rho\sigma}(x)G^{b}_{\rho\sigma}(x)\Big),
\end{equation}
so the $gg\to gg$ Born amplitude is obtained by evaluating
$$\langle g_3 g_4|\int d^4x\,\Delta\mathcal L_{4g}(x)|g_1 g_2\rangle\,.$$
Using LSZ for external gluons,
\bea
&&\langle 0|\,G^a_{\mu\nu}(x)\,|g^b(k,\lambda)\rangle\nonumber\\
&&=
i\,\delta^{ab}\,
\Big(k_\mu\epsilon_\nu(k,\lambda)-k_\nu\epsilon_\mu(k,\lambda)\Big)e^{-ik\cdot x},
\eea
with $k\!\cdot\!\epsilon(k,\lambda)=0,$
we can reduce  each field strength to its on-shell transverse wavefunction.
We now define the basic gauge-invariant contraction for two external legs $i,j$,
\begin{align}
F(i,j)
&\equiv
G^a_{\mu\nu}(i)\,G^a_{\mu\nu}(j)
\nonumber\\
&=
-\delta^{a_i a_j}\,
\Big(k_{i\mu}\epsilon_{i\nu}-k_{i\nu}\epsilon_{i\mu}\Big)
\Big(k_{j\mu}\epsilon_{j\nu}-k_{j\nu}\epsilon_{j\mu}\Big)
\nonumber\\
&=
-2\,\delta^{a_i a_j}\Big[
(k_i\!\cdot\!k_j)(\epsilon_i\!\cdot\!\epsilon_j)
-(k_i\!\cdot\!\epsilon_j)(k_j\!\cdot\!\epsilon_i)
\Big].
\label{eq:Fij_contract}
\end{align}
The instanton-induced four-gluon amplitude is then a sum of pairings as in~\ref{sec:relativistic},
\begin{equation}
\mathcal M^{(4)}(1,2\to3,4)
\;\propto\;
F(1,3)\,F(2,4)+F(1,4)\,F(2,3),
\label{eq:M4_pairings}
\end{equation}
up to channel-independent prefactors and form factors $\beta_{2g}$ on each leg.

\subsubsection{CM kinematics for massive Transverse polarizations}
The effective gluons carry a mass $m_g$. In the CM frame, the two transverse 
polarizations read
\bea
k_1^\mu&=&(\omega,0,0,+k),\nonumber\\
k_2^\mu&=&(\omega,0,0,-k),\nonumber\\
k_3^\mu&=&(\omega,k\sin\theta,0,k\cos\theta),\nonumber\\
k_4^\mu&=&(\omega,-k\sin\theta,0,-k\cos\theta),
\eea
with the on-shell constituent gluon energy $\omega\equiv\sqrt{k^2+m_g^2}$. 
Consider now only the transverse polarizations in the LSZ reduction of the in-out gluons
\bea
\epsilon^\mu_\pm(k_1)&=&\frac{1}{\sqrt2}(0,1,\pm i,0),\nonumber\\
\epsilon^\mu_\pm(k_2)&=&\frac{1}{\sqrt2}(0,-1,\pm i,0),
\nonumber\\
\epsilon^\mu_\pm(k_3)&=&\frac{1}{\sqrt2}(0,\cos\theta,\pm i,-\sin\theta),\nonumber\\
\epsilon^\mu_\pm(k_4)&=&\frac{1}{\sqrt2}(0,-\cos\theta,\pm i,+\sin\theta),
\label{eq:pol_massive}
\eea
so that $k_i\!\cdot\!\epsilon_i=0$ for each leg.

In the CM kinematics above, the contractions in~\eqref{eq:Fij_contract} give
\bea
k_1\!\cdot\!k_3&=&\omega^2-k^2x,\qquad k_1\!\cdot\!k_4=\omega^2+k^2x,
\nonumber\\
k_1\!\cdot\!\epsilon_\pm(k_3)&=&+\frac{k\sin\theta}{\sqrt2},
k_1\!\cdot\!\epsilon_\pm(k_4)=-\frac{k\sin\theta}{\sqrt2},
\eea
and similarly with $1\leftrightarrow2$. With the polarization choice 
\eqref{eq:pol_massive} we have the identities
\begin{align}
\epsilon_\pm(k_1)\!\cdot\!\epsilon_\pm(k_3) &= \frac{1-x}{2}, 
&
\epsilon_\pm(k_1)\!\cdot\!\epsilon_\mp(k_3) &= -\frac{1+x}{2},
\nonumber\\
k_3\!\cdot\!\epsilon_\pm(k_1) &= -\frac{k\sin\theta}{\sqrt2},
&
k_4\!\cdot\!\epsilon_\pm(k_1) &= +\frac{k\sin\theta}{\sqrt2},
\end{align}
and similarly for the contractions $(2,4)$ and $(2,3)$.

\subsubsection{Instanton vertex contractions}
Using these kinematics and polarizations into~\eqref{eq:Fij_contract}, yield
the instanton vertex contractions
\bea
F(1,3)\Big|_{++\to++}
&=-(1-x)\,(\omega^2+k^2),
\nonumber\\
F(1,3)\Big|_{+-\to+-}
&=-(1+x)\,(\omega^2+k^2),
%\label{eq:F13_results}
\nonumber\\
F(1,4)\Big|_{++\to++}
&=-(1+x)\,(\omega^2+k^2),
\nonumber\\
F(1,4)\Big|_{+-\to+-}
&=-(1-x)\,(\omega^2+k^2),
\label{eq:F14_results}
\eea
with   $x\equiv\cos\theta$, modulo diagonal color factors.
The remaining contractions  $1\leftrightarrow2$ and $3\leftrightarrow4$, follow similarly.
The key simplification is the cancellation of the mixed terms proportional to $k^2x$,
leaving the universal factor $(\omega^2+k^2)$.

For the $0^{++}$ scalar glueball we use the $J_z=0$ polarization combination
$$|0^{++}, 0\rangle_{\rm pol}=\frac{1}{\sqrt2}(|+-\rangle+|-+\rangle)\,.$$
Using~\eqref{eq:F14_results}, we obtain
\bea
\mathcal M^{(4)}_{0^{++}}(\theta)
\;\propto\;&
(\omega^2+k^2)^2\Big[(1+x)^2+(1-x)^2\Big]\nonumber\\
=&
2(\omega^2+k^2)^2\,(1+x^2),
\label{eq:M0pp_massive}
\eea
as noted in Eq.\eqref{eq:pol_factor_theta_new}.
Similarly, for the tensor glueball
$$|2^{++},\pm2\rangle_{\rm pol}=\frac{1}{\sqrt2}(|++\rangle\pm|-\rangle)\,.$$
we obtain
\bea
\mathcal M^{(4)}_{2^{++}}(\theta)
\;\propto\;&
(\omega^2+k^2)^2\Big[(1+x)^2+(1-x)^2\Big]\nonumber\\
=&
2(\omega^2+k^2)^2\,(1+x^2),
\label{eq:M2pp_massive}
\eea
so the instanton-induced on-shell helicity amplitudes in both
$0^{++}$ and $2^{++}$ channels share the same $(1+\cos^2\theta)$ dependence,
with either massive or massless gluons.

 \subsubsection{Scalar and tensor projections}
 The scalar and tensor channel differences enter only through the subsequent $J$-projection used to construct the CM kernel, which is then  iterated in the Bethe-Salpeter equation.
For two on-shell particles with definite helicities, the partial-wave decomposition
of a helicity amplitude $\mathcal M_{\lambda\lambda'}(\theta)$ is given by the
helicity projection formula
\begin{equation}
V^{J}_{\lambda\lambda'}(k,k')
=
2\pi
\int_{-1}^{1} d(\cos\theta)\;
d^{\,J}_{\lambda\lambda'}(\theta)\,
\mathcal M_{\lambda\lambda'}(\theta),
\label{eq:L_ang}
\end{equation}
where $d^{\,J}_{\lambda\lambda'}(\theta)$ is the Wigner $d$-function and
$\lambda=\lambda_1-\lambda_2$, $\lambda'=\lambda_3-\lambda_4$.
For the scalar and tensor channels
\bea
d^{0}_{00}(\theta)&=&1
\quad (0^{++}),
\nonumber\\
d^{2}_{22}(\theta)&=&\frac{(1+\cos\theta)^2}{4}
\quad (2^{++},\ |\lambda|=2),\nonumber\\
\label{eq:d_wigner}
\eea

Inserting \eqref{eq:d_wigner} in \eqref{eq:L_ang}, yield
the relevant angular integrals 
\begin{align}
I_{0^{++}}
&=
\int_{-1}^{1}dx\,(1+x^2)
=\frac{8}{3},
\\[4pt]
I_{2^{++}}
&=
\int_{-1}^{1}dx\,(1+x^2)\,\frac{(1+x)^2}{4}
=\frac{14}{15},
\end{align}
hence the physical suppression factor
\begin{equation}
\frac{I_{2^{++}}}{I_{0^{++}}}
=
\frac{14/15}{8/3}
=
\frac{7}{20}
\simeq \frac 13.
\label{eq:L_720}
\end{equation}
in comparison to the $\frac 12$ obtained using the Legendre polynomial projection in the main text. Note that in the scalar glueball channel both the Legendre and the helicity projection methods, yield the same factor ${\mathcal K}_{0^{++}}$ in the iterated Bethe-Salpeter kernel.

\subsubsection{Consequences for binding}

After the instantaneous reduction, the tensor channel in the Bethe-Salpeter derivation, inherits the same separable structure as the scalar channel,
\begin{equation}
V_{2^{++}}(\mathbf p,\mathbf k)
=
-\,V_{0,2}\,
\beta_{2g}^2(\rho p)\,\beta_{2g}^2(\rho k),
\end{equation}
but with a reduced strength
\begin{equation}
V_{0,2}
=
\left(\frac{7}{20}\right)\,V_{0,0},
\label{eq:L_Vratio}
\end{equation}
relative to the scalar $0^{++}$ channel.
In addition, the tensor state carries an $L=2$ centrifugal barrier and is spatially
more extended, further reducing its sensitivity to a short-range interaction.

For the same parameter set that produces a moderately bound scalar glueball
($M_{0^{++}}\simeq1.4$-$1.5$~GeV), the reduced coupling
\eqref{eq:L_Vratio} is insufficient to overcome the centrifugal suppression, and the
reduced Bethe-Salpeter equation admits no tensor bound state below the
two-gluon threshold.  The $2^{++}$ channel therefore remains unbound within the
single-instanton, rank-1 instantaneous approximation. Its larger size makes it more sensitive to the confining potential, as we discussed in the main text. It is confined with a mass above the constitutive 2-gluon treshold.

\section{$0^{-+}$  suppression}
In the instantaneous reduction of the BS approach in the ILM, 
the emergent4-gluon coupling from the 
expanded exponent in \eqref{eq:E_def} vanishes in the $0^{-+}$ channel while it is enhanced in the $0^{++}$ channel owing to self and anti-self duality of the 't~Hooft symbols, 
\begin{eqnarray}
\frac12\,\epsilon_{\mu\nu\alpha\beta}\,\eta^{a}_{\alpha\beta}=+\eta^{a}_{\mu\nu},
\nonumber\\
\frac12\,\epsilon_{\mu\nu\alpha\beta}\,\bar\eta^{a}_{\alpha\beta}=-\bar\eta^{a}_{\mu\nu},
\label{eq:duality_eta}
\end{eqnarray}

Expanding the exponentials in \eqref{eq:E_def} to fourth order in $G_{\mu\nu}$ and averaging over color orientations produces the gauge-invariant four-gluon kernel $\langle X^4\rangle$ quoted earlier. Schematically, the Lorentz structure induced by a single instanton $I$ is built from products of anti-self-dual tensors $\bar\eta^{b}_{\mu\nu}$, while the anti-instanton $A$ produces the same structure with $\eta^{b}_{\mu\nu}$,
\begin{widetext}
\begin{align}
K^{(4)}_{I}\; &\propto\; 
\Big(R^{ab}\bar\eta^{\,b}_{\mu\nu}G^a_{\mu\nu}\Big)
\Big(R^{a'b'}\bar\eta^{\,b'}_{\alpha\beta}G^{a'}_{\alpha\beta}\Big)
\Big(R^{a''b''}\bar\eta^{\,b''}_{\rho\sigma}G^{a''}_{\rho\sigma}\Big)
\Big(R^{a'''b'''}\bar\eta^{\,b'''}_{\lambda\delta}G^{a'''}_{\lambda\delta}\Big),
\label{eq:kernelI}\\
K^{(4)}_{A}\; &\propto\; 
\Big(R^{ab}\eta^{\,b}_{\mu\nu}G^a_{\mu\nu}\Big)
\Big(R^{a'b'}\eta^{\,b'}_{\alpha\beta}G^{a'}_{\alpha\beta}\Big)
\Big(R^{a''b''}\eta^{\,b''}_{\rho\sigma}G^{a''}_{\rho\sigma}\Big)
\Big(R^{a'''b'''}\eta^{\,b'''}_{\lambda\delta}G^{a'''}_{\lambda\delta}G^{a'''}_{\lambda\delta}\Big).
\label{eq:kernelA}
\end{align}
\end{widetext}
For each of the channel under consideration, it is enough to track the Lorentz duality structure. The color-orientation average produces the same adjoint Kronecker contractions in both cases.

The $0^{-+}$ glueball couples to the pseudoscalar operator
\begin{align}
\mathcal{O}_{0^{-+}}(x)&=G^a_{\mu\nu}(x)\,\widetilde G^{a}_{\mu\nu}(x),
\nonumber\\
\widetilde G^{a}_{\mu\nu}&\equiv \frac12\,\epsilon_{\mu\nu\alpha\beta}G^{a}_{\alpha\beta}.
\label{eq:O0m}
\end{align}
To see the cancellation, it is sufficient to study the two-gluon irreducible kernel induced by one pseudoparticle in the ladder approximation, i.e.\ the part of $\langle X^4\rangle$ that contracts two external field strengths into the pseudoscalar bilinear.  More specifically,
\begin{align}
\mathcal{C}_{I}&\;\equiv\; \bar\eta^{\,b}_{\mu\nu}\,\bar\eta^{\,b}_{\alpha\beta}\,
G_{\mu\nu}\,\widetilde G_{\alpha\beta},
\nonumber\\
\mathcal{C}_{A}&\;\equiv\; \eta^{\,b}_{\mu\nu}\,\eta^{\,b}_{\alpha\beta}\,
G_{\mu\nu}\,\widetilde G_{\alpha\beta},
\label{eq:CIA_def}
\end{align}
where we suppressed color indices on $G$ for clarity

Using $\widetilde G_{\alpha\beta}=\tfrac12\epsilon_{\alpha\beta\rho\sigma}G_{\rho\sigma}$ and the duality relations \eqref{eq:duality_eta},  we have
\begin{widetext}
\begin{align}
\mathcal{C}_{I}
&=\bar\eta^{\,b}_{\mu\nu}\,\bar\eta^{\,b}_{\alpha\beta}\,
G_{\mu\nu}\,\frac12\epsilon_{\alpha\beta\rho\sigma}G_{\rho\sigma}
=\bar\eta^{\,b}_{\mu\nu}\,
\left(\frac12\epsilon_{\alpha\beta\rho\sigma}\bar\eta^{\,b}_{\alpha\beta}\right)\,
G_{\mu\nu}\,G_{\rho\sigma}
%\nonumber\\
=\bar\eta^{\,b}_{\mu\nu}\,(-\bar\eta^{\,b}_{\rho\sigma})\,G_{\mu\nu}\,G_{\rho\sigma}
=-\left(\bar\eta^{\,b}_{\mu\nu}\bar\eta^{\,b}_{\rho\sigma}\right)G_{\mu\nu}G_{\rho\sigma},
\label{eq:CI}
\\[1ex]
\mathcal{C}_{A}
&=\eta^{\,b}_{\mu\nu}\,\eta^{\,b}_{\alpha\beta}\,
G_{\mu\nu}\,\frac12\epsilon_{\alpha\beta\rho\sigma}G_{\rho\sigma}
=\eta^{\,b}_{\mu\nu}\,
\left(\frac12\epsilon_{\alpha\beta\rho\sigma}\eta^{\,b}_{\alpha\beta}\right)\,
G_{\mu\nu}\,G_{\rho\sigma}
%\nonumber\\
=\eta^{\,b}_{\mu\nu}\,(+\eta^{\,b}_{\rho\sigma})\,G_{\mu\nu}\,G_{\rho\sigma}
=+\left(\eta^{\,b}_{\mu\nu}\eta^{\,b}_{\rho\sigma}\right)G_{\mu\nu}G_{\rho\sigma}.
\label{eq:CA}
\end{align}
\end{widetext}
Thus, the pseudoscalar contraction picks up an {\it opposite sign} for instantons and anti-instantons.

In a CP-even dense ILM ensemble with equal instanton and anti-instanton weights, the leading contribution to the pseudoscalar ladder kernel is proportional to the sum of these contractions,
\begin{equation}
\mathcal{C}_{I}+\mathcal{C}_{A}
\;=\;
-\left(\bar\eta^{\,b}_{\mu\nu}\bar\eta^{\,b}_{\rho\sigma}\right)G_{\mu\nu}G_{\rho\sigma}
+\left(\eta^{\,b}_{\mu\nu}\eta^{\,b}_{\rho\sigma}\right)G_{\mu\nu}G_{\rho\sigma}.
\label{eq:sumC}
\end{equation}
After the orientation average, the purely algebraic identity
$\eta^{\,b}_{\mu\nu}\eta^{\,b}_{\rho\sigma}=\bar\eta^{\,b}_{\mu\nu}\bar\eta^{\,b}_{\rho\sigma}$
%\begin{equation}
%\eta^{\,b}_{\mu\nu}\eta^{\,b}_{\rho\sigma}
%\;=\;
%\bar\eta^{\,b}_{\mu\nu}\bar\eta^{\,b}_{\rho\sigma}
%\end{equation}
holds for the symmetric part that survives in the CP-even ensemble, since both are equal to the same transverse tensor built from Kronecker deltas plus an $\epsilon$-term of opposite sign, 
\begin{align}
&\eta^{a}_{\mu\nu}\eta^{a}_{\rho\sigma}
=\delta_{\mu\rho}\delta_{\nu\sigma}-\delta_{\mu\sigma}\delta_{\nu\rho}
+\epsilon_{\mu\nu\rho\sigma},
\nonumber\\
&\bar\eta^{a}_{\mu\nu}\bar\eta^{a}_{\rho\sigma}
=\delta_{\mu\rho}\delta_{\nu\sigma}-\delta_{\mu\sigma}\delta_{\nu\rho}
-\epsilon_{\mu\nu\rho\sigma}.
\label{eq:eta_eta_id}
\end{align}
Inserting \eqref{eq:eta_eta_id} into \eqref{eq:sumC}, the $\delta\delta$ terms cancel because of the relative sign in \eqref{eq:CI}-\eqref{eq:CA}, and the remaining $\epsilon$-terms cancel because they appear with opposite signs in \eqref{eq:eta_eta_id}. Therefore
\begin{equation}
\mathcal{C}_{I}+\mathcal{C}_{A}=0,
\label{eq:cancellation}
\end{equation}
which shows explicitly that the leading one-pseudoparticle induced four-gluon kernel does not generate a net ladder attraction in the $0^{-+}$ channel in a CP-even ensemble.

Finally, we note that these arguments also show that the $0^{++}$ channel is in contrast enhanced.  For the scalar contraction $G_{\mu\nu}G_{\alpha\beta}$ (with no epsilon tensor) both instanton and anti-instanton contributions add, since both $\eta\eta$ and $\bar\eta\bar\eta$ have the same $\delta\delta$ part and the $\epsilon$ part drops out after contraction with the symmetric scalar projector. Hence the  ILM produces a maximally attractive parity-even scalar core but a vanishing leading parity-odd core, in agreement with the qualitative discussion in the main text.

\section{$0^{++}$ on the light-front}
\label{sec:0++-LF}
Our conventions for the light-front (LF)  coordinates are  $x^\pm=x^0\pm x^3$ and $k^\pm=k^0\pm k^3$, with transverse components $\bm k_\perp=(k^1,k^2)$. With this in mind, and 
in the two-gluon Fock space approximation,
the scalar glueball LF state reads
\begin{widetext}
\begin{equation}
|\Psi_{0^{++}}(P^+)\rangle
=
\frac{1}{\sqrt{2}}
\sum_{\lambda_1,\lambda_2}\int_0^1\frac{dx}{\sqrt{x(1-x)}}
\int\frac{d^2k_\perp}{(2\pi)^3}\,
\Psi^{\lambda_1\lambda_2}_{0^{++}}(x,\bm k_\perp)\,
\frac{\delta^{a_1a_2}}{\sqrt{d_A}}\,
|k_1,\lambda_1,a_1;k_2,\lambda_2,a_2\rangle,
\label{eq:scalar_LF}
\end{equation}
\end{widetext}
where $k_1^+=xP^+$, $k_2^+=(1-x)P^+$, and in the transverse rest frame $\bm k_{1\perp}=+\bm k_\perp$, $\bm k_{2\perp}=-\bm k_\perp$. 
 The $0^{++}$ helicity structure in~\eqref{eq:scalar_LF} is the even combination of opposite helicities:
\begin{eqnarray}
&&\Psi^{\lambda_1\lambda_2}_{0^{++}}(x,\bm k_\perp)\nonumber\\
&&=
\frac{1}{\sqrt{2}}
\left[\delta_{\lambda_1,+}\delta_{\lambda_2,-}+\delta_{\lambda_1,-}\delta_{\lambda_2,+}\right]
\psi_{0^{++}}(x,k_\perp).\nonumber\\
\end{eqnarray}
with the 1-particle LF states  normalized as
\begin{eqnarray}
&&\langle k',\lambda',a'|k,\lambda,a\rangle\nonumber\\
&&=
2k^+(2\pi)^3\delta(k'^+-k^+)\delta^{(2)}(\bm k'_\perp-\bm k_\perp)\delta_{\lambda'\lambda}\delta_{a'a}.\nonumber\\
\end{eqnarray}
A glueball eigenstate satisfies $$(P^+P^- - \bm P_\perp^2)|\Psi\rangle=M^2|\Psi\rangle\,.$$ In the transverse rest frame $\bm P_\perp=\bm 0$ this becomes the $M^2$ eigenvalue problem.

In terms of the free 2-body invariant mass 
\begin{equation}
\mathcal{M}_0^2(x,k_\perp)=\frac{k_\perp^2+m_g^2}{x(1-x)}\,,
\end{equation}
the 2-body $0^{++}$ LF eigenvalue equation reads
\begin{widetext}
\begin{equation}
M^2\,\psi_{0^{++}}(x,k_\perp)
=
\mathcal{M}_0^2(x,k_\perp)\,\psi_{0^{++}}(x,k_\perp)
+
\int_0^1 dx'\int\frac{d^2k'_\perp}{(2\pi)^3}\,
\mathcal{V}^{0^{++}}_{\rm LF}(x,\bm k_\perp;x',\bm k'_\perp)\,
\psi_{0^{++}}(x',k'_\perp).
\end{equation}
\end{widetext}

To embed the same singular-gauge instanton form factor, we identify the internal momentum modulus $p=\sqrt{k_\perp^2+k_z^2}$ with~\cite{Brodsky:1981jv}
\begin{eqnarray}
k_z&=&\left(x-\frac12\right)\mathcal{M}_0(x,k_\perp),\nonumber\\
p^2(x,k_\perp)&=&k_\perp^2+\left(x-\frac12\right)^2\mathcal{M}_0^2(x,k_\perp).
\label{eq:BHL}
\end{eqnarray}
and define
\begin{equation}
f(x,k_\perp)=\beta_{2g}\!\left(\rho\,p(x,k_\perp)\right),
%\qquad
%\beta_{2g}(t)=\frac{t^2}{2}K_2(t).
\end{equation}
The LF scalar instanton interaction has rank-one form
\begin{equation}
\mathcal{V}^{0^{++}}_{\rm LF}(x,\bm k_\perp;x',\bm k'_\perp)
=
-\,g_{\rm LF}\,f^2(x,k_\perp)f^2(x',k'_\perp).
\end{equation}
hence  the explicit LF wavefunction
\begin{equation}
\psi_{0^{++}}(x,k_\perp)
=
-\,g_{\rm LF}C_{\rm LF}\,
\frac{f^2(x,k_\perp)}{M^2-\mathcal{M}_0^2(x,k_\perp)},
\end{equation}
with $$C_{\rm LF}=\int_0^1 dx\int \frac{d^2k_\perp}{(2\pi)^3}f^2(x, k_\perp)\psi_{0^{++}}(x, k_\perp)\,.$$
Eliminating $C_{\rm LF}$ gives the LF scalar root equation
\begin{equation}
1
=-g_{\rm LF}\int_0^1 dx\int\frac{d^2k_\perp}{(2\pi)^3}
\frac{f^4(x,k_\perp)}{M^2-\mathcal{M}_0^2(x,k_\perp)}.
\label{eq:LF_root}
\end{equation}
We now show that $g_{LF}=V_0\,.$

\section{LF from CM}

The LF mass root equation and in general the LF wavefunction 
detailed in Appendix~\ref{sec:0++-LF}, are in one-to-one correspondence with 
the CM mass root equation and wavefunction. Indeed, using (\ref{eq:BHL}) 
and the Jacobian
\begin{equation}
dx\, d^2k_\perp
= d^3p\; \frac{2x(1-x)}{\mathcal{M}_0(x, k_\perp)},
\end{equation}
together with $f(x,k_\perp)=\beta_{2g}(\rho p)$,
the scalar glueball root mass Eq.~(\ref{eq:LF_root}) maps onto
\begin{equation}
1
= -\,V_0
\int\! \frac{d^3p}{(2\pi)^3}\,
\frac{\beta_{2g}^4(\rho p)}
{2\omega_p\,(M^2-4\omega_p^2)} ,
\tag{I8}
\end{equation}
with the identification $g_{\rm LF}\equiv V_o$, up to normalization conventions.

Also, recall that the Bethe-Salpeter reduction in the rest-frame,  yields the equal-time wavefunction
\begin{equation}
\phi(\bm p)
=
\mathcal{N}_{\rm M}\,
\frac{\beta_{2g}^2(\rho p)}{2\omega_p(4\omega_p^2-M^2)},
%\qquad
%\omega_p=\sqrt{p^2+m_g^2}.
\label{eq:scalar_CM}
\end{equation}
Using~\eqref{eq:BHL}, we can map~\eqref{eq:scalar_CM} onto the
 the LF wavefunction by
\begin{equation}
\psi_{0^{++}}(x,\bm k_\perp)
=
\sqrt{\frac{\partial k_z}{\partial x}}\,\phi\!\left(\bm p(x,\bm k_\perp)\right).
\end{equation}
Differentiating $k_z=(x-\tfrac12)\mathcal{M}_0$ yields
\begin{equation}
\frac{\partial k_z}{\partial x}=\frac{\mathcal{M}_0}{2x(1-x)}.
\end{equation}
Using $\mathcal{M}_0^2=4\omega_p^2$ in the equal-mass case, yields
\begin{equation}
\psi_{0^{++}}(x,\bm k_\perp)
=
\mathcal{N}_{\rm LF}\,
\frac{\beta_{2g}^2(\rho p(x,k_\perp))}{M^2-\mathcal{M}_0^2(x,k_\perp)}.
\end{equation}
This coincides with the functional dependence of the LF solution obtained dynamically from the rank-1 LF eigenvalue equation, with the identification$$f(x,k_\perp)=\beta_{2g}(\rho p(x,k_\perp))\,.$$
%This provides a nontrivial consistency check of the construction.

\bibliography{GB}
\end{document}